\documentclass[twocolumn,preprintnumbers,amsmath,amssymb]{revtex4}
\UseRawInputEncoding
\usepackage{amsmath}
\usepackage{extarrows}
\usepackage{graphicx}
\usepackage{bm}
\usepackage[all]{xy}
\usepackage{amsmath}
\usepackage{multirow}
\usepackage{bbm}
\usepackage{euscript}
\usepackage{amssymb}
\usepackage{extarrows}
\usepackage{bbm}
\usepackage{graphicx}
\usepackage{epstopdf}
\usepackage{color,xcolor}

\newtheorem{theorem}{Theorem}
\newtheorem{definition}{Definition}

\newtheorem{lemma}{Lemma}

\begin{document}
\title{General spin systems without genuinely multipartite nonlocality}

\author{Yan-Han Yang, Xue Yang, Ming-Xing Luo}

\affiliation{\small{} School of Information Science and Technology, Southwest Jiaotong University, Chengdu 610031, China}

\begin{abstract}
There are multipartite entangled states in many-body systems which may be potential resources in various quantum applications. There are lots of methods to witness specific entangled systems. However, no efficient method is available to explore many-body systems without multipartite entanglement. It may provide necessary restrictions for experimental preparations of multipartite entanglement. Our goal is to solve this problem for spin systems. The maximal effective velocity with propagation of information is bounded in quantum spin systems with short-range interactions from Lieb-Robinson's inequalities. This implies two clustering theorems for ground states and thermal states. With these propagation relations, we show that both the gapped ground state and thermal state at an upper-bounded inverse temperature have no genuine multipartite nonlocality when disjoint regions are far away from each other. The present $n$-particle system shows only biseparable quantum correlations when the propagation relations show exponential decay. Similar result holds for spin systems with product states as initial states. These results show interesting features of quantum many-body systems with exponential decay of correlations.
\end{abstract}
\maketitle

\section{Introduction}

Quantum systems may have strong correlations which can not be generated from any classical systems. This is originated from entanglement that is a remarkable feature in quantum mechanics \cite{Schrodinger}. It implies that the global state of special composite systems  cannot be written into a product of single states or its mixture \cite{HHH}. So far, the quantum entanglement has become the central concept in quantum information processing \cite{BCM} with prospective applications such as quantum computation \cite{Nielsen,Lago-Rivera}, quantum cryptography \cite{Ekert,Chen}, quantum teleportation \cite{Bennett1,li1}, dense coding \cite{Bennett2,Galindo}, quantum radar \cite{Malik}, entanglement swapping \cite{Zukowski1,Wang}, and remote states preparation \cite{Babichev}.

Until now, photons may be the most promising candidates for experimentally encoded entanglement because of its robustness against environments \cite{Edamatsu1}. Parametric down-conversion in optical nonlinear crystals is widely used to prepare entangled photons \cite{Edamatsu1,Yao} with semiconductor sources \cite{Edamatsu1,Edamatsu2}. These kind of entangled systems can be applied for long-distance quantum communication \cite{XLWang} and quantum networks \cite{JPL}. For quantum computations \cite{Nielsen,Lago-Rivera}, quantum information processing \cite{Sorensen,Prevedel}, and quantum metrology beyond the standard quantum limit \cite{Feldmann}, it may be desirable to involve as many particles as possible into entangled states \cite{Xu}. In this case, the large-scale entanglement can be in principle built by the system evolution of many-body systems such as interacted spin systems associated with special Hamiltonian \cite{Xu,XYLuo}.

To verify an experimentally entangled system, Bell shows \cite{Bell} that the quantum correlations derived from local measurements can be stronger than those from any classical physics or hidden variable models. This is further extended to the CHSH inequality \cite{Clauser} for general bipartite systems or noisy states \cite{Werner}. Interestingly, different from bipartite states, multipartite states allow different local models. One is from the bipartite nonlocality \cite{GHZ,Zukowski2} which cannot guarantee the existence of fully nonlocal correlations between $n$-particle. The other is for ruling out separable states or biseparable entangled states. The so-called genuinely multipartite nonlocality can be verified by Svetlichny inequality \cite{Svetlichny}. This is further extended for multiple settings \cite{Collins,Bancal1} or the Hardy inequality \cite{Hardy}. Another stronger nonlocality is named as the network entanglement which do not allow network decompositions \cite{Kimble,net1,Navascues,Kraft,Luo2020}. Besides, there are lots of results related to multipartite nonlocality \cite{Batle,Campbell,Wagner,Pelisson} and nonlocality in translationally invariant spin Hamiltonians \cite{Tura}.

Generally, it is difficult to characterize all multipartite entangled states. Various special cases have been considered in many-body systems \cite{Hofmann,Biswas,Jindal,Roy,Szalay}. Specific conditions are used to verify the nonlocality in lattice models \cite{Sun,Pelisson,Dhar}. In quantum many-body systems, the ground state has been proved to be a highly entangled system \cite{Giampaolo1}. The entanglement witness is applied in spin chains \cite{HHH,Guhne1} based on the ground state energy. This can be extended for the ground and thermal states of spin-1/2 Hamiltonians in any lattice geometry \cite{Stasinska}. These results are focused on sufficient conditions for verifying multipartite entanglement in experiments. On one hand, there are few results related to sufficient conditions of many-body systems for preparing multipartite entanglement in experiment. This is very important for experiments and applications. On the other hand, there is no general result with which one cannot generate multipartite entanglement in experiment. As a necessary condition, this is crucial for settings an experiment. Our motivation in this work is to find necessary condition for spin systems in lattice. The main idea is inspired by the exponentially clustering correlations \cite{Vieira}. We show that any local measurements acting on remote regions of given quantum spin systems cannot create genuinely multipartite nonlocal correlations in terms of any Svetlichny-type inequality in the biseparable model \cite{Svetlichny}.  Similar result holds for product states as the initial states. These results provide new insights in experimental preparations of multipartite entanglement with lattice systems.

The rest is organized as follows: Sec.II briefly describes biseparable quantum correlations and exponential clustering correlations for many-body systems. Sec.III contributes the ground state of the gap Hamiltonian, that is, it does not show genuinely multipartite nonlocality. The biseparability of thermal equilibrium states is shown in Sec.IV. Sec.V devotes to the product states as the initial states while the last section concludes the paper.

\section{Preliminaries}

\subsection{Biseparable quantum correlations}

Consider a tripartite scenario. One source distributes three physical systems to distant observers, Alice, Bob, and Charlie. Each observer measures its system under local measurement basis. The measurement chosen by Alice is labeled $x$ and results in output $a$. Similarly, Bob chooses measurement $y$ and gets outcome $b$. Charlie chooses measurement $z$ and gets outcome $c$. The outcome is characterized by the joint probability distribution $p(abc|xyz)$. For a given correlation $p(abc|xyz)$, it is local \cite{Bell} if $p(abc|xyz)$ can be decomposed into:
\begin{eqnarray}
p(abc|xyz)=\int_{\Omega}q(\lambda)p(a|x,\lambda)p(b|y,\lambda) p(c|z,\lambda)d\mu(\lambda)
\label{eq01}
\end{eqnarray}
where $(\Omega, q(\lambda), \mu(\lambda))$ is the measure space of the hidden variable $\lambda$, which accounts for the dependence among $a$, $b$, and $c$ \cite{Brunner,Almeida}.
The nonlocal correlations can not be decomposed into Eq.(\ref{eq01}). For quantum scenarios, the nonlocal correlations can only be obtained from tripartite entangled state, which is not fully separable \cite{Bell}. The nonlocality of bipartite quantum state can be verified by using Bell inequality \cite{Bell,Clauser}.

Svetlichny \cite{Svetlichny} proposed a new kind of multipartite nonlocality if a tripartite correlations cannot always be decomposed into
\begin{eqnarray}
p(abc|xyz)&=
&\int_{\Omega_1}q(\lambda_1)p(ab|xy,\lambda_1)p(c|z,\lambda_1)d\mu(\lambda_1)
\nonumber
\\
&&+\int_{\Omega_2}q(\lambda_2)p(ac|xz,\lambda_2) p(b|y,\lambda_2)d\mu(\lambda_2)
\nonumber\\
\nonumber &&+\int_{\Omega_3}q(\lambda_3)p(bc|yz,\lambda_3) p(a|x,\lambda_3)d\mu(\lambda_3)
\\
\label{eq02}
\end{eqnarray}
where $(\Omega_i,q(\lambda_i),\mu(\lambda_i))$ denotes the measure space of the hidden variable $\lambda_i$, $\int_{\Omega_i}q(\lambda_i)=1$ and $q(\lambda_i)\geq0$, $i=1, 2, 3$. The correlations shown in Eq.(\ref{eq02}) is named as biseparable correlations which can be easily followed from the local measurements on the biseparable states given by
\begin{eqnarray}
\rho=p_{1}\rho_{AB}\otimes\rho_{C}+p_{2}\rho_{AC}\otimes \rho_{B} + p_{3} \rho_{BC}\otimes \rho_{A}
\label{eq02a}
\end{eqnarray}
where $\{ p_{1},p_{2},p_{3}\}$ is the probability distribution over different decompositions, $\rho_{A(B,C)}$ denotes the states of the system $A$, $B$ or $C$, and $\rho_{AB(BC,AC)}$ denotes the state of the joint system $A$ and $B$, $B$ and $C$, or $A$ and $C$. These kind of correlations are not completely separable. This provides a new method to verify so-called genuinely multipartite nonlocality by using Svetlichny inequality \cite{Svetlichny,Bancal2}. This can be extended for $n$-partite states. An genuinely $n$-partite entangled cannot be decomposed into
\begin{eqnarray}
\rho= \sum_{I_{1},\cdots,I_{k}} p_{I_{1}\cdots{}I_{k}} \rho_{I_{1}\cdots{}I_{k}}
\label{eq2.1}
\end{eqnarray}
where $\{ p_{I_{1}\cdots{}I_{k}} \}$ is a probability distribution over all possible $k$-partite partition $I_1,\cdots,I_k$, and $\rho_{I_{1}\cdots{}I_{k}}$ denotes to the $k$-partite separable states for $k$-partite partition $I_1,\cdots,I_k$, $k=2,\cdots, n$.

\subsection{Polytopes of biseparable correlations}

Similar to fully separable states \cite{Bell}, all the biseparable states consist of convex set. It means that the genuinely multipartite nonlocality can be defined as the convex combination of a finite number of extremal points with given finite inputs and outputs \cite{Brunner}. This allows to verify the genuinely multipartite nonlocality by using linear Bell inequalities, that is, facet inequalities. Take the tripartite correlations shown in Eq.(\ref{eq02}) as an example. The first solution to characterise this polytope with dichotomic inputs and outputs is Svetlichny inequality \cite{Svetlichny} given by
\begin{eqnarray}
 S_{SI}&:=\nonumber&\langle A_{0}B_{0}C_{0}\rangle+\langle A_{1}B_{0}C_{0}\rangle+\langle A_{0}B_{1}C_{0}\rangle
 \\ \nonumber&&+\langle A_{0}B_{0}C_{1}\rangle-\langle A_{0}B_{1}C_{1}\rangle-\langle A_{1}B_{1}C_{0}\rangle
 \\ \nonumber&&-\langle A_{1}B_{0}C_{1}\rangle-\langle A_{1}B_{1}C_{1}\rangle
 \\ &\leq&4
\label{eq31}
\end{eqnarray}
which holds for any biseparable state $\rho$, where $\langle XYZ\rangle={\rm tr}(X\otimes{}Y\otimes{}Z\rho)$. The maximal violation of quantum correlations is $4\sqrt{2}$. A general Bell-bilocality inequality can be written as:
\begin{eqnarray}
  \sum_{k_{1},k_{2},k_{3}}  \psi_{k_{1}k_{2}k_{3}}\langle E_{k_{1}}^{(1)}E_{k_{2}}^{(2)}E_{k_{3}}^{(3)}\rangle_{\rho} \leq\Delta_{loc}
\label{eq3.1}
\end{eqnarray}
which holds for any biseparable states $\rho$ in Eq.(\ref{eq02a}), where $\psi_{k_{1}k_{2}k_{3}}$ are constants, $\{E_{k_{i}}^{(i)}\}$ denotes the observable of the party $i$ conditional on input $k_i$, $i=1, 2, 3$, and $\Delta_{loc}$ denotes the upper bound in terms of any biseparable quantum correlations.

Consider an $n$-partite Bell experiment, in which each party $i$ measures its own system under the observables $\{E^{(i)}_{k_i}\}$. To characterise the polytope of genuinely multipartite nonlocal correlations, we define a general $n$-partite Bell inequality involved multipartite correlators $\langle E^{(i_{1})}_{k_{1}}\cdots E^{(i_{s})}_{k_{s}}\rangle$ as
\begin{eqnarray}
\sum^{n}_{s=1}\sum^{n}_{i_{1}\neq\cdots\neq i_{s}=1}\sum_{k_{1},\cdots,k_{s}}
\psi^{(i_{1}\cdots{}i_{s})}_{k_{1}\cdots{}k_{s}}\langle E^{(i_{1})}_{k_{1}}\cdots E^{(i_{s})}_{k_{s}}\rangle \leq \Delta_{loc}
\label{eq40}
\end{eqnarray}
which holds for any states $\rho$ in Eq.(\ref{eq2.1}), where $\psi^{(i_{1}\cdots{}i_{s})}_{k_{1}\cdots{}k_{s}}$ are constants, $\langle E^{(i_{1})}_{k_{1}}, \cdots, E^{(i_{s})}_{k_{s}}\rangle_{\rho}={\rm tr}(E^{(i_{1})}_{k_{1}}\otimes\cdots {}\otimes{}E^{(i_{s})}_{k_{s}}\otimes{}\mathbbm{1}_{\overline{i_1\cdots{}i_s}}\rho)$ with the identity operator on the complement system ($\overline{i_1\cdots{}i_s}$) of $i_1,\cdots{},i_s$, and $\Delta_{loc}$ is the upper bound in terms of quantum states in Eq.(\ref{eq2.1}).

\begin{definition}
For any disjoint regions in a lattice of the quantum spin system, given a real number $\varepsilon>0$, the total state of the $n$-partite system is presented by $\rho$ acting on a finite dimensional Hilbert space ${\cal H}_{1}\otimes\cdots \otimes {\cal H}_n$. $\rho$ is $\varepsilon$-local in terms of the genuinely $m$-partite nonlocality respect to these disjoint regions if the following inequality holds
\begin{eqnarray}
 {\cal S}(\rho)\leq \Delta_{loc}+\varepsilon
\label{eqdef}
\end{eqnarray}
where $\Delta_{loc}$ is defined in the inequality (\ref{eq40}).

\end{definition}

From Definition 1, any local measurements acting on a $\varepsilon$-local state $\rho$ cannot create genuinely multipartite nonlocal correlations  \cite{Svetlichny}. Our considerations here are any facet inequalities for verifying the genuinely multipartite nonlocality in the biseparable model \cite{Svetlichny}. For simplicity, we assume that all observables satisfy $\|E^{(i)}_{k_i}\|\leq 1$ \cite{Nielsen}. Otherwise, we can define $E^{(i)}_{k_i}=E^{(i)}_{k_i}/c$ with $c=\max\{\|E^{(i)}_{k_i}\|\}$.

\subsection{Quantum spins systems and clustering theorems}

For a given quantum spin system, it is schematically represented by a lattice (a finite set of vertices) $\Omega$ equipped with a metric $d$, which represents the distance between the sits in the lattice \cite{John}. Each vertex $x\in\Omega$ is associated with a finite dimensional Hilbert space ${\cal H}_\Omega$ in sense that for each $X\subset\Omega$ the associated Hilbert space is the tensor product ${\cal H}_{X}=\otimes_{x\in X}{\cal H}_{x}$. The observable in $X$ is denoted by ${\cal B}({\cal H}_{X})$. The bounded linear operator over ${\cal H}_X$ for the support of $\{A\in{\cal B}({\cal H}_{\Omega})\mid A=A_{X}\otimes I_{\Omega/X}\}$ is given by the smallest set of $X\subset\Omega$ \cite{John}. An interaction for such a system is a mapping $h_{\Omega}$ from the set of subsets of $\Omega$ to ${\cal B}({\cal H}_{\Omega})$, that is $h_{X}$ has support in $X$. The Hamiltonian is defined by $H_{\Omega}=\Sigma_{X\subset\Omega}h_{X}$. The dynamics of the model is defined by $A(t)=e^{itH}Ae^{-itH}$. Finally, as in the general construction of finite quantum spin systems, we assume that $\tau$ is the distance of the interaction. With the additional assumption, all the involved interactions are short-range, i.e., the distance is small with respect to the size of the lattice.

\textit{Example 1}. Consider finite interval $[1,L]$ of arbitrary length $L\in \mathbb{N}$ with a $\frac{1}{2}$-spin at each site. The Hilbert space is given by ${\cal H}_{[1,L]}=\otimes_{j=1}^L\mathbb{C}^2$.  The Hamiltonian of an isotropic XY chain is given by
\begin{eqnarray}
{\cal H}_{[1,L]}=\sum_{j=1}^{L-1}
h_{j,j+1}+\sum_{j=1}^Lt_{j,s}
\end{eqnarray}
which consists of next-neighbor interactions $h_{j,j+1}=-\sigma_j^X\sigma_{j+1}^X\
-\sigma_j^Y\sigma_{j+1}^Y$ between the sites $j$ and $j+1$ with $j=1, \cdots, L-1$. $t_{j}=c_j\sigma_j^Z$ denotes local field acting at site $j$, $j=1,\cdots,L$. Here, $\sigma^X,\sigma^Y$ and $\sigma^Z$ are Pauli matrices. Another example is anisotropic XY chain define by next-neighbor interactions
$h_{j,j+1}=-(1+\gamma)\sigma_j^X\sigma_{j+1}^X\
-(1-\gamma)\sigma_j^Y\sigma_{j+1}^Y$ with an anisotropy parameter $\gamma\not=0$.

\begin{figure}
\begin{center}
\resizebox{160pt}{150pt}{\includegraphics{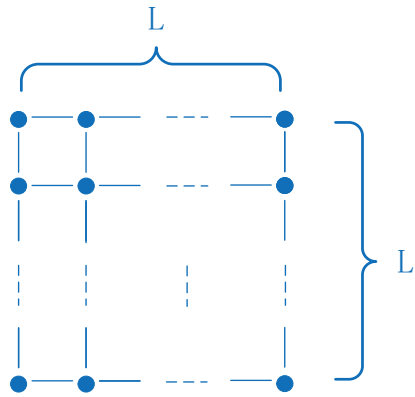}}
\label{FIG1}\caption{Finite square of $[1,L]\times [1,L]$.}
\end{center}
\end{figure}

\textit{Example 2}. Consider finite squares $[1,L]\times [1,L]$ of arbitrary length $L\in \mathbb{N}$ with a $\frac{1}{2}$-spin at each site, as shown Fig.1. The Hilbert space is given by ${\cal H}_{[1,L]\times [1,L]}=\otimes_{j,s}^L\mathbb{C}^2$. Suppose that the Hamiltonian is given by
\begin{eqnarray}
{\cal H}_{[1,L]\times [1,L]}=\sum_{d((j_1,s_1),(j_2,s_2))=1}
h_{(j_1,s_1),(j_2,s_2)}
+\sum_{j,s=1}^Lt_{j,s}
\end{eqnarray}
which consists of next-neighbor interactions $h_{(j_1,s_1),(j_2,s_2)}$ between the sites $(j_1,s_1)$ and $(j_2,s_2)$ with $d((j_1,s_1),(j_2,s_2))=|j_1-j_2|+|s_1-s_2|=1$, $j=1, \cdots, L-1$. $t_{j,s}$ denotes local field acting at site $j$, $j=1,\cdots,L$.

\textit{Example 3}. Consider infinite squares $\mathbb{Z}\times \mathbb{Z}$ with a $\frac{1}{2}$-spin at each site. Suppose that the Hamiltonian is given by
\begin{eqnarray}
{\cal H}_{\mathbb{Z}\times \mathbb{Z}}=\sum_{d((j_1,s_1),(j_2,s_2))\leq k}h_{(j_1,s_1),(j_2,s_2)}
+\sum_{j,s\in \mathbb{Z}}t_{j,s}
\end{eqnarray}
which consists of $k$-neighbor interactions $h_{(j_1,s_1),(j_2,s_2)}$ between the sites $(j_1,s_1)$ and $(j_2,s_2)$ with $d((j_1,s_1),(j_2,s_2))=|j_1-j_2|+|s_1-s_2|\leq k$, $j_1,j_2,s_1,s_2\in\mathbb{Z}$,  and $t_{j,s}$ denotes local field acting at site $j$, $j\in\mathbb{Z}$. $k$ is the interaction distance.

\textit{Example 4}. Consider $m$-dimensional cubic $V_m\subset \mathbb{Z}^{\times m}$ with a $\frac{1}{2}$-spin or Heisenberg anti-ferromagnet at each site. Suppose that the Hamiltonian is given by
\begin{eqnarray}
{\cal H}_{V_m}=\sum_{x,y\in V_m, \atop{ d(x,y)\leq k}}h_{x,y}
+\sum_{x\in V_m}t_{x}
\end{eqnarray}
which consists of $k$-neighbor interactions $h_{x,y}$ between the sites $x$ and $y$ with $d(x,y)=\sum_{j=1}^m|x_j-y_j|\leq k$, and $t_{x}$ denotes local field acting at site $x$, $x, y\in V_m$.

For these systems with specific assumptions, there are some states with exponentially clustering correlations. We start with the ground state of a gapped Hamiltonian \cite{Hastings,Nachtergaele2}. The distance between two regions is expressed as the minimum distance between vertices in the two regions respectively: $d_{X,Y}=\min_{a\in X,b\in Y}\{d(a,b)\}$. In the following, the minimal distance between any two regions of $X$, $Y$ and $Z$ is lower bounded by $\tau$, i.e.,
\begin{eqnarray}
\label{Eq00}
d_{X,Y}, d_{Y,Z}, d_{X, Z}\geq \tau
\end{eqnarray}
for any $X,Y,Z\subset\Omega$. Without lose of generality, assume that $|X|=\max\{|X|,|Y|,|Z|\}$.

\begin{lemma}\cite{Hastings,Nachtergaele2}
For a system with the ground state $\rho$ and a spectral gap $\Delta E>0$ above the ground-energy, there exist constants $c$, $\kappa >0$ such that
\begin{eqnarray}
|\langle ABC\rangle_{\rho}-\langle A\rangle_{\rho}\langle BC\rangle_{\rho}|\leq ce^{-\kappa\tau}\|A\|\,\|B\|\,\|C\|\,|X|
\end{eqnarray}
where $A$, $B$, $C\in{\cal B}({\cal H}_{\Omega})$ are observables supported in the disjoint regions $X$, $Y$ and $Z$ in the lattice, respectively.

\label{01}
\end{lemma}

In Lemma 1, $\langle A\rangle_{\rho}$ denotes the expect of the observables $A, \mathbbm{1}_{B}, \mathbbm{1}_{C}$ with the identity operator $\mathbbm{1}$, and $\langle BC\rangle_{\rho}$ denotes the expect of the observables $\mathbbm{1}_{A}, B, C$. $c$ and $\tau$ in Lemma 1 are related to the lattice geometry, the maximal interaction energy and the spectral gap. It is independent of the regions $X$ ,$Y$ or $Z$. Here, $X$ and $Y$ are disjoint regions mean that they are no common vertex.

The thermal state is another kind of states with exponentially clustering correlations \cite{Kliesch}. Especially, a thermal state, or Gibbs state associated with a Hamiltonian $H$ at inverse temperature $\beta$ is given by
\begin{eqnarray}
\rho(\beta)=\frac{e^{-\beta H}}{\Phi(\beta)}
\label{eq05}
\end{eqnarray}
with the partition function $\Phi(\beta)={\rm Tr}(e^{-\beta H})$. There is a universal inverse critical temperature $\beta^{*}$ such that the correlation decays exponentially for $\beta<\beta^{*}$. Here, $\beta^{*}$ is independent of the system size. It depends on the typical energy of interaction and the spatial dimension of the lattice.

\begin{lemma}
\cite{Kliesch}
If $\rho(\beta)$ is a thermal state at an inverse temperature $\beta<\beta^{*}$, there exist constants $c$, $\kappa >0$ such that
\begin{eqnarray}
\nonumber
|\langle ABC\rangle_{\rho_{(\beta)}}-\langle A\rangle_{\rho_{(\beta)}}\langle BC\rangle_{\rho_{(\beta)}}| \leq & c(\beta)e^{-\kappa(\beta)\tau} \|A\|\,\|B\|\,\|C\|
\\
\end{eqnarray}
where $A$, $B$, $C\in{\cal B}({\cal H}_{\Omega})$ are observables supported in the disjoint regions $X$, $Y$ and $Z$ in the lattice, respectively.

\label{02}
\end{lemma}

The proofs of Lemmas 1 and 2 depend on the Lieb-Robinson's bounds \cite{Fredenhagen,Lieb}, which show that the maximal effective velocity with propagation of information is bounded in quantum spin systems with short-range interactions. The Lieb-Robinson bound is also applied for featuring how much correlation can be created by using a product state \cite{Nachtergaele3}. It increases exponentially over time, but decreases exponentially with distance \cite{Nachtergaele3}.

\begin{lemma} \cite{Nachtergaele3}
Let $\rho(0)=\otimes_{x\in\Omega}\rho_{x}$ be the initial state of the lattice.  We have
\begin{eqnarray}
\nonumber|\langle ABC\rangle_{\rho_{(t)}}-\langle A\rangle_{\rho_{(t)}}\langle BC\rangle_{\rho_{(t)}}|
&\leq&
ce^{-\kappa\tau}(e^{\kappa v t}-1) \|A\|\,\|B\|\,
\\
& &\times{}\|C\|\,|X|\,|Y|\,|Z|
\end{eqnarray}
where $A$, $B$, $C\in{\cal B}({\cal H}_{\Omega})$ are observables supported in the disjoint regions $X$, $Y$ and $Z$ in the lattice, respectively.

\label{03}
\end{lemma}

\section{Gapped ground states}

In this section, we discuss quantum correlations in multipartite spin systems by using the clustering theorems of gapped Hamiltonian.

\subsection{Tripartite systems}

\begin{figure}
\begin{center}
\resizebox{160pt}{100pt}{\includegraphics{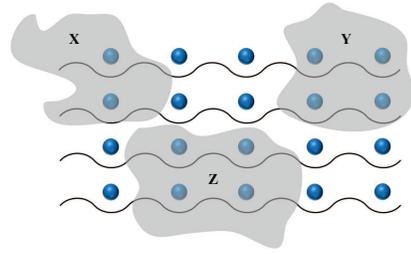}}
\caption{Tripartite scenario in spin lattice.}
\label{FIG2}
\end{center}
\end{figure}

Consider an experiment via spin systems defined on the lattice $V$. Let $\Omega$ be a finite set of spins in the lattice $V$, where spins interact with each other by a given Hamiltonian operator. The interaction is short-range relative to the size of the lattice. The Hilbert space associated with $\Omega$ is defined by ${\cal H}_\Omega$. There are observables $E_{k_{1}}^{(1)}$, $E_{k_{2}}^{(2)}$ and $E_{k_{3}}^{(3)}$ acting on the lattice supported in the regions $X$, $Y$ and $Z$ ($\{X, Y, Z\}\subset\Omega$), respectively, as shown in Fig.2. In this tripartite system, three spatially separated agents share a quantum state and each party $i$ measure its own system under the measurements $E_{k_{1}}^{(1)}$, $E_{k_{2}}^{(2)}$ and $E_{k_{3}}^{(3)}$. $|I|$ is the number of sites in the region $I\subset\Omega$. The total state is represented by $\rho$. We define the following quantity:
\begin{eqnarray}
{\cal S}^{X,Y,Z}(\rho)=\sup_{E_{k_{1}}^{(1)},E_{k_{2}}^{(2)},E_{k_{3}}^{(3)}}{\cal S}^{X,Y,Z}(\rho,E_{k_{1}}^{(1)},E_{k_{2}}^{(2)},E_{k_{3}}^{(3)})
\label{eq3.10}
\end{eqnarray}
where $ {\cal S}^{X,Y,Z}(\rho,E_{k_{1}}^{(1)},E_{k_{2}}^{(2)},E_{k_{3}}^{(3)})$ is defined by according to Bell operators from Eq.(\ref{eq3.1}) as
\begin{eqnarray}
 \nonumber &&{\cal S}^{X,Y,Z}(\rho,E_{k_{1}}^{(1)},E_{k_{2}}^{(2)},E_{k_{3}}^{(3)})
  \\
  &=&\sum_{k_{1},k_{2},k_{3}}  \psi_{k_{1}k_{2}k_{3}}\langle E_{k_{1}}^{(1)}E_{k_{2}}^{(2)}E_{k_{3}}^{(3)}\rangle_{\rho}
\label{eq3.2}
\end{eqnarray}
The goal in what follows is to show that the violation of the inequality (\ref{eq3.1}) is vanishingly small when three regions $X$, $Y$ and $Z$ are far apart from each other.

The $\varepsilon$-bilocal state means that it has no genuinely tripartite nonlocality approximately, that is, quantum correlations of the specific system are biseparable for all facet inequalities (\ref{eq3.1}). This is associated with the given regions $X$, $Y$ and $Z$, as shown in Fig.2. We have $\varepsilon\ll 1$ such the tripartite system behaves in local in the biseparable model \cite{Svetlichny}.

Consider a tripartite ground state $\rho$ associated with a gapped Hamiltonian. In general, we show that the genuinely tripartite entangled correlations should be very small for distant parts when $\rho$ shows exponentially clustering of correlations. Informally, we prove that a given quantum correlation is biseparable when the involved regions are far apart. Let $\eta$ be a constant defined by:
\begin{eqnarray}
\eta =\sum_{k_{1},k_{2},k_{3}}|
\,\psi_{k_{1}k_{2}k_{3}}|
\label{eq3.3}
\end{eqnarray}

\begin{theorem}
\label{04}
Let $\rho$ be the tripartite ground state of a gapped Hamiltonian. There exist constants $c$, $\kappa>0$ such that $\rho$ is $\varepsilon$-local with respect to any disjoint regions $X$, $Y$ and $Z\subseteq\Omega$, where $\varepsilon=\eta ce^{-\kappa\tau}|X|$.

\end{theorem}

\textbf{Proof}. From Lemma \ref{01} we get that
\begin{eqnarray}
\psi_{k_{1}k_{2}k_{3}}\langle E_{k_{1}}^{(1)}E_{k_{2}}^{(2)}E_{k_{3}}^{(3)}\rangle
&\leq& \psi_{k_{1}k_{2}k_{3}} \langle E^{(1)}_{k_{1}}\rangle\langle E^{(2)}_{k_{2}} E^{(3)}_{k_{3}}\rangle
\nonumber
\\
&&
+ce^{-\kappa\tau}  |X| \, |\psi_{k_{1}k_{2}k_{3}}|
\label{eq3.4}
\end{eqnarray}
Combining with Eq.(\ref{eq3.2}), we get that
\begin{eqnarray}
\nonumber
&&{\cal S}^{XYZ}(\rho,E_{k_{1}}^{(1)},E_{k_{2}}^{(2)},E_{k_{3}}^{(3)})
\\ \nonumber
&\leq &\sum_{k_{1},k_{2},k_{3}}  (\psi_{k_{1}k_{2}k_{3}} \langle E^{(1)}_{k_{1}}\rangle\langle E^{(2)}_{k_{2}} E^{(3)}_{k_{3}}\rangle
\nonumber\\
&& +ce^{-\kappa\tau} |X| \, |\psi_{k_{1}k_{2}k_{3}}|)
\label{eq3.5}
\end{eqnarray}
We define $\hat{\cal S}^{XYZ}(\rho,E_{k_{1}}^{(1)},E_{k_{2}}^{(2)},E_{k_{3}}^{(3)})$ as the expect of Bell operators with correlations of almost two bodies, i.e.,
\begin{eqnarray}
\nonumber && \hat{\cal S}^{XYZ}(\rho,E_{k_{1}}^{(1)},E_{k_{2}}^{(2)},E_{k_{3}}^{(3)})
\\
 &= & \sum_{k_{1},k_{2},k_{3}}  \psi_{k_{1}k_{2}k_{3}} \langle E^{(1)}_{k_{1}}\rangle\langle E^{(2)}_{k_{2}} E^{(3)}_{k_{3}}\rangle
\label{eq3.6}
\end{eqnarray}
From Eqs.(\ref{eq3.3}) and (\ref{eq3.6}) we get that
\begin{eqnarray}
{\cal S}^{XYZ}(\rho,E_{k_{1}}^{(1)},E_{k_{2}}^{(2)},E_{k_{3}}^{(3)})  \leq\Delta_{bi}+\eta ce^{-\kappa\tau}|X|
\label{eq3.7}
\end{eqnarray}
According to the inequality (\ref{eq3.7}), the upper bound of ${\cal S}^{X,Y,Z}(\rho,E_{k_{1}}^{(1)},E_{k_{2}}^{(2)},E_{k_{3}}^{(3)})$ does not depend on the measurements of $E_{k_{1}}^{(1)},E_{k_{2}}^{(2)},E_{k_{3}}^{(3)}$. Hence, it implies the following inequality
\begin{eqnarray}
{\cal S}^{XYZ}(\rho)
&=&\sup_{E_{k_{1}}^{(1)},E_{k_{2}}^{(2)},E_{k_{3}}^{(3)}}{\cal S}^{XYZ}(\rho,E_{k_{1}}^{(1)},E_{k_{2}}^{(2)},E_{k_{3}}^{(3)})
\nonumber\\
&\leq&
\Delta_{bi}+\eta ce^{-\kappa\tau}|X|
\label{eq3.8}
\end{eqnarray}
This completes the proof. $\hfill\square$

\textit{Example 5}. Svetlichny inequality \cite{Svetlichny} is useful for characterizing the nonlocality of three-body systems. For each run of the experiment, one of two available dichotomy measurements are performed: $A_{i}, B_{j}, C_{k}$ on each site respectively, $i,j,k\in \{0,1\}$. The violation of Svetlichny inequality (\ref{eq31}) shows the genuinely tripartite nonlocality. We define the following quantity:
\begin{eqnarray}
{\cal S}^{XYZ}_{SI}(\rho)=\sup_{A,B,C}{\cal S}^{XYZ}_{SI}(\rho,A,B,C)
\label{eq32}
\end{eqnarray}
where $ {\cal S}^{XYZ}_{SI}(\rho,A,B,C)$ is defined by according to Svetlichny operators as
\begin{eqnarray}
 {\cal S}^{XYZ}_{SI}(\rho,A,B,C)
  &=&\langle A_{0}B_{0}C_{0}\rangle_{\rho}+\langle A_{1}B_{0}C_{0}\rangle_{\rho}
 \nonumber \\
 &&+\langle A_{0}B_{1}C_{0}\rangle_{\rho}+\langle A_{0}B_{0}C_{1}\rangle_{\rho}
  \nonumber\\
 &&-\langle A_{0}B_{1}C_{1}\rangle_{\rho}-\langle A_{1}B_{1}C_{0}\rangle_{\rho}
  \nonumber\\
 \nonumber
 &&-\langle A_{1}B_{0}C_{1}\rangle_{\rho}-\langle A_{1}B_{1}C_{1}\rangle_{\rho}
 \\
\label{eq33}
\end{eqnarray}
According to Lemma \ref{04}, for each set of measurement operators $A_{i}$, $B_{j}$ and $C_{k}$, we have
\begin{eqnarray}
\nonumber |\langle A_{i}B_{j}C_{k}\rangle-\langle A_{i}\rangle\langle B_{j}C_{k}\rangle|\leq ce^{-\kappa\tau} \|A_{i}\|\,\|B_{j}\|\,\|C_{k}\|\,|X|
\\
\label{eq34}
\end{eqnarray}
From the assumptions of $\|A_{i}\|, \|B_{j}\|, \|C_{k}\|\leq1$, we get
\begin{eqnarray}
\langle A_{i}B_{j}C_{k}\rangle\leq\langle A_{i}\rangle\langle B_{j}C_{k}\rangle+ce^{-\kappa\tau}|X|
\label{eq35}
\end{eqnarray}
From the assumptions of $\langle A_{i}\rangle\leq1$ and $\langle B_{j}C_{k}\rangle\leq1$ with $i,j,k\in\{0,1\}$, combined with Eq.(\ref{eq33}) and the inequality (\ref{eq35}) we have
\begin{eqnarray}
 \nonumber&&{\cal S}^{XYZ}_{SI}(\rho,A,B,C)
 \\ \nonumber& \leq&  \langle A_{0}\rangle(\langle B_{0}C_{0}\rangle+\langle B_{1}C_{0}\rangle+\langle B_{0}C_{1}\rangle-\langle B_{1}C_{1}\rangle)
 \\ \nonumber&&+\langle A_{1}\rangle(\langle B_{0}C_{0}\rangle-\langle B_{1}C_{0}\rangle-\langle B_{0}C_{1}\rangle-\langle B_{1}C_{1}\rangle)
 \\ \nonumber&& +8 ce^{-\kappa\tau}|X|
 \\ \nonumber&\leq&|\langle B_{0}C_{0}\rangle+\langle B_{1}C_{0}\rangle+\langle B_{0}C_{1}\rangle-\langle B_{1}C_{1}\rangle|
 \\ \nonumber&&+|\langle B_{0}C_{0}\rangle-\langle B_{1}C_{0}\rangle-\langle B_{0}C_{1}\rangle-\langle B_{1}C_{1}\rangle|+8 ce^{-\kappa\tau}|X|
 \\ \nonumber&\leq &2\max\{|\langle B_{0}C_{0}\rangle|+|\langle B_{1}C_{1}\rangle|,\,|\langle B_{0}C_{1}\rangle|+|\langle B_{1}C_{0}\rangle|\}
 \\ \nonumber&& +8 ce^{-\kappa\tau}|X|
 \\ &\leq &4+8 ce^{-\kappa\tau}|X|
\label{eq38}
\end{eqnarray}
Hence, From the inequality (\ref{eq38}) we get that
\begin{eqnarray}
{\cal S}^{XYZ}_{SI}(\rho,A,B,C) \leq  4+8 ce^{-\kappa\tau}|X|
\label{eq39}
\end{eqnarray}
According to the inequality (\ref{eq39}), $\rho$ is $\varepsilon$-bilocal for Svetlichny inequality \cite{Svetlichny} with respect to disjoint regions of $X$, $Y$ and $Z$, where $\varepsilon=8 ce^{-\kappa\tau}|X|$.

Here, we want to underline that $c$ and $\kappa$ are independent of the regions. The main reason is that the three parties in the system are far apart and $\tau$ is large. Hence,  $\varepsilon$ is very small.

For one-dimensional cluster-Ising model with $\frac{1}{2}$-spin, whether the ground state satisfies or breaks the symmetries of the Hamiltonian has no effect on the the amount of genuine tripartite entanglement \cite{Giampaolo1}. Moreover, if the observable is acted on distant regions of the lattice, the ground state of gapped Hamiltonian behave almost locally by clustering the bipartite system into two uncorrelated particles \cite{Vieira}. Combing with Lemma \ref{04}, we show that the ground state of gapped Hamiltonian is unable to significantly violate any Bell inequality (\ref{eq3.1}) when the involved regions are far apart, that is, there is almost no genuinely tripartite nonlocality for the ground state.

\subsection{$n$-partite systems}

Consider a scenario with an $n$-partite Bell experiment, in which each party measure its own system under the measurements $E_{k_i}^{(i)}$. The genuinely $n$-partite nonlocality can be verified by using the inequality (\ref{eq40}). As shown in Fig.3, the action of each agent is restricted to disjoint regions. The measurement from the $i$-th agent is an operator on the lattice, supported by the region $X_{i}$. Similar to Eq.(\ref{eq3.10}), we define
\begin{equation}
  {\cal S}^{X_{1}\cdots{}X_{n}}(\rho)=\sup_{E^{(1)},\cdots, E^{(n)}}{\cal S}^{X_{1}\cdots{}X_{n}}(\rho,E^{(1)},\cdots, E^{(n)}),
 \label{eq41}
\end{equation}
where ${\cal S}^{X_{1}\cdots{}X_{n}}(\rho,E^{(1)},\cdots{},E^{(n)})$ is given by
\begin{eqnarray}
  &&{\cal S}^{X_{1}\cdots{}X_{n}}(\rho,E^{(1)},\cdots{},E^{(n)})
 \nonumber \\
 \nonumber &=&\sum^{n}_{s=1}\sum^{n}_{i_{1}\neq\cdots\neq i_{s}=1}\sum_{k_{1},\cdots,k_{s}}\psi^{(i_{1}\cdots{}i_{s})}_{k_{1}\cdots{}k_{s}}\langle E^{(i_{1})}_{k_{1}}\cdots E^{(i_{s})}_{k_{s}}\rangle_{\rho}
\\
 \label{eq42}
\end{eqnarray}
For any state $\rho$, it shows the genuinely $n$-partite nonlocality if its correlations derived from local measurements violate the inequality (\ref{eq40}), that is, ${\cal S}^{X_{1}\cdots{}X_{n}}(\rho)>\Delta_{bi}$.

\begin{figure}
\begin{center}
\resizebox{180pt}{120pt}{\includegraphics{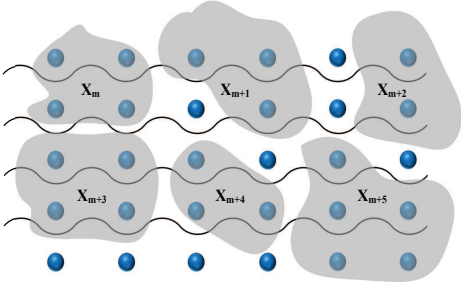}}
\caption{$n$-partite scenario in a spin lattice.}
\label{FIG3}
\end{center}
\end{figure}

We firstly extend exponentially clustering of correlations from the assumptions in Lemma \ref{01}.

\textbf{Fact 1.} \textit{For any set of disjoint regions $X_{1},\cdots,X_{s}\subset\Omega$, if a ground state $\rho$ acting on ${\cal H}_\Omega$ shows exponentially clustering of correlations, then for any set of operators $E_{1},\cdots,E_{s}$ supported in $X_{1},\cdots,X_{s}$ we have
\begin{eqnarray}
  &&|\langle E_{1}\cdots{} E_{s}\rangle_{\rho}-\langle E_{1}\rangle_{\rho}\cdots\langle E_{s-2}\rangle_{\rho}\langle E_{s-1}E_{s}\rangle_{\rho}|
 \nonumber \\
&\leq&(s-2)ce^{-\kappa\tau} \prod_{t=1}^s\|E_{t}\| \,|X|
\end{eqnarray}
where $c$, $\kappa$ are defined on exponentially clustering of correlations, $\tau=\min\{\tau_{ij}\}$, $\tau_{ij}$ is the distance between the regions $X_{i}$ and $X_{j}$, and $|X|=\max\{|X_{1}|,\cdots,|X_{s}|\}$.}

The proof of Fact 1 is shown in Appendix A. Let $\gamma$ be a constant defined by:
\begin{eqnarray}
\gamma=\sum^{n}_{s=1}\sum^{n}_{i_{1}\neq\cdots\neq i_{s}=1}\sum_{k_{1},\cdots,k_{s}}(s-2)|
\,\psi^{(i_{1}\cdots{}i_{s})}_{k_{1}\cdots{}k_{s}}|
\label{eq49}
\end{eqnarray}

\begin{theorem}\label{07}
If $\rho$ is the ground state of a gapped Hamiltonian, there exist $c$, $\kappa>0$. Such for any disjoint regions $X_{1},\cdots{},X_{n}\subset\Omega$, we have $\rho$ is $\varepsilon$-local with $\varepsilon = \gamma ce^{-\kappa\tau}|X|$.

\end{theorem}

\textbf{Proof}. From Fact 1 and Lemma \ref{01} we get
\begin{eqnarray}
\nonumber&&\psi^{(i_{1}\cdots{}i_{s})}_{k_{1}\cdots{}k_{s}}\langle E^{(i_{1})}_{k_{1}}\cdots{} E^{(i_{s})}_{k_{s}}\rangle
\\ \nonumber&\leq&\psi^{(i_{1}\cdots{}i_{s})}_{k_{1}\cdots{}k_{s}}\langle E^{(i_{1})}_{k_{1}}\rangle\cdots\langle E^{(i_{s-2})}_{k_{s-2}}\rangle\langle E^{(i_{s-1})}_{k_{s-1}} E^{(i_{s})}_{k_{s}}\rangle
\\ && +(s-2)ce^{-\kappa\tau} |\psi^{(i_{1}\cdots{}i_{s})}_{k_{1}\cdots{}k_{s}}| \,|X|
\label{eq50}
\end{eqnarray}
From Eq.(\ref{eq42}) we get that
\begin{eqnarray}
\nonumber
&&{\cal S}^{X_{1}\cdots{}X_{n}}(\rho,E^{(1)},\cdots,E^{(n)})
\\ \nonumber
&\leq &\sum^{n}_{s=1}\sum^{n}_{i_{1}\neq\cdots\neq i_{s}=1}\sum_{k_{1},\cdots,k_{s}}(\psi^{(i_{1}\cdots{}i_{s})}_{k_{1}\cdots{}k_{s}}  \langle E^{(i_{1})}_{k_{1}}\rangle\cdots\langle E^{(i_{s-2})}_{k_{s-2}}\rangle
\nonumber\\
&& \times\langle E^{(i_{s-1})}_{k_{s-1}} E^{(i_{s})}_{k_{s}}\rangle
+(s-2)ce^{-\kappa\tau}|X|\, |\psi^{(i_{1}\cdots{}i_{s})}_{k_{1}\cdots{}k_{s}}|)
\label{eq51}
\end{eqnarray}

Define $\hat{\cal S}^{X_{1}\cdots{}X_{n}}(\rho,E^{(1)},\cdots, E^{(n)})$ as the expect of Bell operator based on any system with correlations of almost two bodies, that is,
\begin{eqnarray}
\nonumber&&\hat{\cal S}^{X_{1}\cdots{}X_{n}}(\rho,E^{(1)},\cdots,E^{(n)})
\\
&=&\sum^{n}_{s=1}\sum^{n}_{i_{1}\neq\cdots\neq i_{s}=1}\sum_{k_{1},\cdots,k_{s}}\psi^{(i_{1}\cdots{}i_{s})}_{k_{1}\cdots{}k_{s}}
\langle E^{(i_{1})}_{k_{1}}\rangle\cdots\langle E^{(i_{s-2})}_{k_{s-2}}\rangle
\nonumber\\
&& \times \langle E^{(i_{s-1})}_{k_{s-1}} E^{(i_{s})}_{k_{s}}\rangle
\label{eq52}
\end{eqnarray}
From the inequality (\ref{eq40}) and Eqs.(\ref{eq49}) and (\ref{eq52}) we get that
\begin{eqnarray}
{\cal S}^{X_{1}\cdots{} X_{n}}(\rho,E^{(1)},\cdots, E^{(n)})\leq\Delta_{bi}+\gamma ce^{-\kappa\tau} |X|
\label{eq53}
\end{eqnarray}
where $\Delta_{bi}$ is defined in the inequality (\ref{eq40}).

According to the inequality (\ref{eq53}), the upper bound of ${\cal S}^{X_{1}\cdots{}X_{n}}(\rho,E^{(1)},\cdots,E^{(n)})$ does not depend on the measurements of $E^{(1)},\ldots,E^{(n)}$. Hence, it implies the following inequality
\begin{eqnarray}
{\cal S}^{X_{1}\cdots{}X_{n}}(\rho)\leq\Delta_{bi}+\gamma ce^{-\kappa\tau} |X|
\label{eq55}
\end{eqnarray}
where $\Delta_{bi}$ is defined in the inequality (\ref{eq40}). This completes the proof. $\hfill\square$

\textit{Example 6}. Seevinck and Svetlichny derive $n$-particle Bell-type inequalities under the assumption of partial separability \cite{Seevinck}. These are direct generalizations of the tripartite inequalities \cite{Svetlichny} as
\begin{eqnarray}
|S_{n}^{\pm}|= | \sum_{I}\nu_{t(I)}^{\pm} E_{k_{1}}^{(1)}\cdots E_{k_{n}}^{(n)} | \leq 2^{n-1}
\label{eq3.20}
\end{eqnarray}
where $I=(k_{1},\cdots,k_{n})$, $t(I)$ is the number that index 1 appears in $I$, and $\nu_{k}^{\pm}=(-1)^{k(k\pm1)/2}$ \cite{Seevinck}. Define the following two equations:
\begin{eqnarray}
&CH_{1}=\langle A_{0}B_{0}\rangle+\langle A_{0}B_{1}\rangle+\langle A_{1}B_{0}\rangle-\langle A_{1}B_{1}\rangle
\label{eq3.21}
\\
&CH_{2}=\langle A_{0}B_{0}\rangle-\langle A_{0}B_{1}\rangle-\langle A_{1}B_{0}\rangle-\langle A_{1}B_{1}\rangle
\label{eq3.22}
\end{eqnarray}
Combing the inequality (\ref{eq50}) with Eqs.(\ref{eq3.20}), (\ref{eq3.21}) and (\ref{eq3.22}), we get
\begin{eqnarray}
\nonumber |S_{n}^{\pm}| & \leq&
2^{n-2}|CH_{1}|+ 2^{n-2}|CH_{2}| +(n-2)2^{n}ce^{-\kappa\tau}|X|
\\ \nonumber
& \leq& 2^{n-2} \max\{ |CH_{1}+CH_{2}|, |CH_{1}-CH_{2}| \}
\\ \nonumber
& &+(n-2)2^{n}ce^{-\kappa\tau}|X|
\\
& \leq& 2^{n-1} +(n-2)2^{n}ce^{-\kappa\tau}|X|
\label{eq3.23}
\end{eqnarray}

So far, we have generalized Lemma \ref{01} to the $n$-particle systems. Note that there are two partitions for a four-partite system from Eq.(\ref{eq2.1}). This inspires another clustering method for the genuinely multipartite nonlocality of lattice systems.

\begin{lemma}\label{30}
Given four disjoint regions $X$, $Y$, $Z$ and $U\subset\Omega$, a system with the ground state $\rho$ and a spectral gap $\Delta E>0$ above the ground-energy, there exist constants $c$ and $\kappa >0$ such that
\begin{eqnarray}
|\langle ABCD\rangle_{\rho}-\langle AB\rangle_{\rho}\langle CD\rangle_{\rho}|&\leq & ce^{-\kappa\tau} \|A\|\,\|B\|
\nonumber\\
&&\times\,\|C\|\,\|D\|\,|X|
\label{eqlemma31}
\\
|\langle ABCD\rangle_{\rho}-\langle AC\rangle_{\rho}\langle BD\rangle_{\rho}|&\leq & ce^{-\kappa\tau} \|A\|\,\|B\|
\nonumber\\
&&\times\,\|C\|\,\|D\|\,|X|
\label{eqlemma32}
\\
|\langle ABCD\rangle_{\rho}-\langle AD\rangle_{\rho}\langle BC\rangle_{\rho}|&\leq & ce^{-\kappa\tau} \|A\|\,\|B\|
\nonumber\\
&&\times\,\|C\|\,\|D\|\,|X|
\label{eqlemma33}
\end{eqnarray}
where $A$, $B$, $C$ and $D\in{\cal B}({\cal H}_{\Omega})$ are observables supported in the regions $X$, $Y$, $Z$ and $U$, respectively.
\end{lemma}

\textbf{Proof}. As $\rho$ is the ground state of a gapped Hamiltonian, we can obtain from Lemma \ref{01} that
\begin{eqnarray}
|\langle ABC\rangle-\langle A\rangle\langle BC\rangle|  \leq ce^{-\kappa\tau'}|X'|
\label{eqlemma3.21}
\end{eqnarray}
where $\tau'$ is the minimal distance between any two regions of $X$, $Y$ and $Z$, $|X'|=\max\{|X|,|Y|,|Z|\}$. We take $X$ and $Y$ as one region, $Z$ and $U$ as respectively separate regions. From Lemma \ref{01} we get
\begin{eqnarray}
|\langle ABCD\rangle-\langle AB\rangle\langle CD\rangle|  \leq  ce^{-\kappa\tau}|X|
\label{eqlemma3.22}
\end{eqnarray}
where $\tau$ is the minimal distance between any two regions of $X$, $Y$, $Z$ and $U$, $|X|=\max\{|X|,|Y|,|Z|,|U|\}$. The same procedure may be easily applied to obtain Eqs.(\ref{eqlemma32}) and (\ref{eqlemma33}). This completes the proof. $\square$

\textbf{Fact 2.} \emph{For any set of disjoint regions $X_{1},\cdots,X_{s}\subset\Omega$, if a ground state $\rho$ acting on ${\cal H}_\Omega$ shows exponentially clustering of correlations with $c$, $\kappa>0$, we have
\begin{eqnarray}
  \nonumber &&|\langle E_{1}\cdots{} E_{s}\rangle_{\rho}-\langle E_{i_{1}}\cdots{} E_{i_{k}}\rangle_{\rho}\langle E_{i_{k+1}}\cdots{} E_{i_{s}}\rangle_{\rho}|
  \\ \nonumber&\leq&
  ce^{-\kappa\tau} \prod_{i=1}^s\|E_{i}\| \,|X|
\end{eqnarray}
where $E_{1},\cdots, E_{s}$ are observables supported in $X_{1},\cdots,X_{s}$, respectively.
}

\textbf{Proof}. For the bipartition of $\hat{X}_{1}=\{X_{i_{1}},\cdots,X_{i_{k}}\}$ and $\hat{X}_{2}=\{X_{i_{k+1}},\cdots,X_{i_{s}}\}$,
we have $d(\hat{X}_{1},\hat{X}_{2}) =\min \{ {d(X_{i},X_{j})} \}$ with $i\in \{ i_{1},\ldots,i_{k} \}$ and $j\in \{ i_{k+1},\ldots, i_s \}$. For any $i$ and $j$, we obtain $d(\hat{X}_{1},\hat{X}_{2}) \leq d(X_{i},X_{j}) \leq \tau$. It means that the minimal distance between $\cup_{i=i_{1}}^{i_{k}} X_{i}$ and  $\cup_{i=i_{k+1}}^{i_{s}} X_{i}$ is no less than $\tau$. So, according to Lemma \ref{30}, we get that
\begin{eqnarray}
  \nonumber &&|\langle E_{1}\cdots{} E_{s}\rangle_{\rho}-\langle E_{i_{1}}\cdots{} E_{i_{k}}\rangle_{\rho}\langle E_{i_{k+1}}\cdots{} E_{i_{s}}\rangle_{\rho}|
  \\&\leq& ce^{-\kappa\tau}  \prod_{i=1}^s\|E_{i}\|\,|X|
\label{eq66}
\end{eqnarray}
This completes the proof. $\square$

Fact 2 is useful for the clustering theorems. For simplicity, define $\hat{\gamma}$ as
\begin{eqnarray}
\hat{\gamma}=\sum^{n}_{s=1}\sum^{n}_{i_{1}\neq\cdots\neq i_{s}=1}\sum_{k_{1},\cdots,k_{s}}
|\psi^{(i_{1}\cdots{}i_{s})}_{k_{1}\cdots{}k_{s}}|
\label{eq68}
\end{eqnarray}

\begin{theorem}\label{12}
Suppose that $\rho$ is the ground state of a gapped Hamiltonian. There exist $c$, $\kappa>0$ such that $\rho$ is $\varepsilon$-local with respect to any disjoint regions $X_{1},\cdot\cdot\cdot,X_{n}\subset\Omega$, where $\varepsilon= \hat{\gamma}ce^{-\kappa\tau}|X|$.

\end{theorem}

\textbf{Proof}. From Lemma \ref{30} and Fact 2 we get that
\begin{eqnarray}
\nonumber&&\psi^{(i_{1}\cdots{}i_{s})}_{k_{1}\cdots{} k_{s}}\langle E^{(i_{1})}_{k_{1}}\cdots{} E^{(i_{s})}_{k_{s}}\rangle
\\ \nonumber&\leq&\psi^{(i_{1}\cdots{}i_{s})}_{k_{1}\cdots{}k_{s}}\langle E^{(i_{1})}_{k_{1}}\cdots{} E^{(i_{m})}_{k_{m}}\rangle\langle E^{(i_{m+1})}_{k_{m+1}}\cdots{} E^{(i_{s})}_{k_{s}}\rangle
\\ &&+ce^{-\kappa\tau} |X|\, |\psi^{(i_{1}\cdots{}i_{s})}_{k_{1}\cdots{}k_{s}}|
\label{eq69}
\end{eqnarray}
Combining with Eq.(\ref{eq42}), it follows that
\begin{eqnarray}
\nonumber&&{\cal S}^{X_{1}\cdots{}X_{n}}(\rho,E^{(1)},\cdots,E^{(n)})
\\ \nonumber&\leq &
\sum^{n}_{s=1}\sum^{n}_{i_{1}\neq\cdots{}\neq i_{s}=1}\sum_{k_{1},\cdots{},k_{s}} \psi^{(i_{1}\cdots{}i_{s})}_{k_{1}\cdots{}k_{s}}\langle E^{(i_{1})}_{k_{1}}\cdots{} E^{(i_{m})}_{k_{m}}\rangle
\nonumber
\\
&& \times \langle E^{(i_{m+1})}_{k_{m+1}}\cdots E^{(i_{s})}_{k_{s}}\rangle
\nonumber\\
&&+ce^{-\kappa\tau}|X|\sum^{n}_{s=1}\sum^{n}_{i_{1}\neq\cdots\neq i_{s}=1}\sum_{k_{1},\cdots,k_{s}}
|\psi^{(i_{1}\cdots{}i_{s})}_{k_{1}\cdots{}k_{s}}|
\label{eq70}
\end{eqnarray}
Similar to the proof of Theorem \ref{07}, define the following quantity:
\begin{eqnarray}
\nonumber&&\tilde{\cal S}^{X_{1}\cdots{}X_{n}}(\rho,E^{(1)}\cdots{}E^{(n)})
\\
&=&\sum^{n}_{s=1}\sum^{n}_{i_{1}\neq\cdots{}\neq i_{s}=1}\sum_{k_{1},\cdots{},k_{s}} \psi^{(i_{1}\cdots{}i_{s})}_{k_{1}\cdots{}k_{s}}\langle E^{(i_{1})}_{k_{1}}\cdots E^{(i_{m})}_{k_{m}}\rangle
\nonumber\\
&&\times \langle E^{(i_{m+1})}_{k_{m+1}}\cdots E^{(i_{s})}_{k_{s}}\rangle
\label{eq71}
\end{eqnarray}
which is associated with an $n$-partite system $\rho$ defined in Eq.(\ref{eq2.1}). From the inequality (\ref{eq40}) we get that
\begin{eqnarray}
\nonumber
&&{\cal S}^{X_{1}\cdots{}X_{n}}(\rho,E^{(1)},\cdots,E^{(n)})
\\ \nonumber
&\leq&
 \tilde{\cal S}^{X_{1}\cdots{}X_{n}}(\rho,E^{(1)},\cdots,E^{(n)})
\\ \nonumber
 && +ce^{-\kappa\tau}|X|\sum^{n}_{s=1}\sum^{n}_{i_{1}\neq\cdots\neq i_{s}=1}\sum_{k_{1},\cdots,k_{s}} |\psi^{(i_{1}\cdots{}{}i_{s})}_{k_{1}\cdots{}k_{s}}|
\\
&\leq&\Delta_{bi}+\hat{\gamma}ce^{-\kappa\tau}|X|
\label{eq72}
\end{eqnarray}
where $\Delta_{bi}$ is defined in the inequality (\ref{eq40}).

According to the inequality (\ref{eq72}), the upper bound of ${\cal S}^{X_{1}\cdots{}X_{n}}(\rho)$ is independent of $E^{(1)},\cdots,E^{(n)}$. It follows that
\begin{eqnarray}
{\cal S}^{X_{1}\cdots{}X_{n}}(\rho)&=&\sup_{E^{(1)},\cdots,E^{(n)}}{\cal S}^{X_{1}\cdots{}X_{n}}(\rho,E^{(1)},\cdots,E^{(n)})
\nonumber
\\
&\leq&\Delta_{bi}+\hat{\gamma}ce^{-\kappa\tau}|X|
\label{eq74}
\end{eqnarray}
where $\Delta_{bi}$ is defined in the inequality (\ref{eq40}). This completes the proof. $\square$

\textit{Example 7}. Svetlichny inequality \cite{Seevinck} for a four-particle system is shown as follows:
\begin{eqnarray}
\nonumber S_{4}&=&\langle A_{0}B_{0}C_{0}D_{0}\rangle-\langle A_{1}B_{0}C_{0}D_{0}\rangle-\langle A_{0}B_{1}C_{0}D_{0}\rangle
\\ \nonumber&&-\langle A_{0}B_{0}C_{1}D_{0}\rangle-\langle A_{0}B_{0}C_{0}D_{1}\rangle-\langle A_{1}B_{1}C_{0}D_{0}\rangle
\\ \nonumber&&-\langle A_{1}B_{0}C_{1}D_{0}\rangle-\langle A_{1}B_{0}C_{0}D_{1}\rangle-\langle A_{0}B_{1}C_{1}D_{0}\rangle
\\ \nonumber&&-\langle A_{0}B_{1}C_{0}D_{1}\rangle-\langle A_{0}B_{0}C_{1}D_{1}\rangle+\langle A_{1}B_{1}C_{1}D_{0}\rangle
\\ \nonumber&&+\langle A_{1}B_{1}C_{0}D_{1}\rangle+\langle A_{1}B_{0}C_{1}D_{1}\rangle+\langle A_{0}B_{1}C_{1}D_{1}\rangle
\\ \nonumber&&+\langle A_{1}B_{1}C_{1}D_{1}\rangle
\\ &\leq &8
\label{eq56}
\end{eqnarray}
Assume that each of four partes has two measurement operators acting on the lattice with support in the region $X$, $Y$, $Z$ and $U$ respectively, where $X,Y,Z,U\subset\Omega$. Note that $X$, $Y$, $Z$ and $U$ are far apart. It means that $\varepsilon$ is very small. In this case, the system cannot show the genuinely four-partite nonlocality. From Lemma \ref{30}, for each set of measurement operators $A_{i}$, $B_{j}$, $C_{k}$ and $D_{l}$, we get that
\begin{eqnarray}
&&
|\langle A_{i}B_{j}C_{k}D_{l}\rangle-\langle A_{i}B_{j}\rangle\langle C_{k}D_{l}\rangle|
\nonumber\\
&\leq & ce^{-\kappa\tau} \|A_{i}\|\,\|B_{j}\|\,\|C_{k}\|\,\|D_{l}\|\,|X|
\label{eq59}
\end{eqnarray}
As in the proof of Fact 1, $\langle A_{i}B_{j}\rangle$ has the spectrum in $[-1,1]$, and $\|A_{i}\|, \|B_{j}\|, \|C_{k}\|, \|D_{l}\|\leq 1$. This implies that
\begin{eqnarray}
\langle A_{i}B_{j}C_{k}D_{l}\rangle\leq\langle A_{i}B_{j}\rangle\langle C_{k}D_{l}\rangle+ce^{-\kappa\tau}|X|
\label{eq60}
\end{eqnarray}
Define four-partite Svetlichny quantity of ${\cal S}^{X,Y,Z,U}_{4}(\rho)$ as follows:
\begin{eqnarray}
{\cal S}^{X,Y,Z,U}_{4}(\rho)=\mathrm{sup}_{A,B,C,D}{\cal S}^{X,Y,Z,U}_{4}(\rho,A,B,C,D),
\label{eq57}
\end{eqnarray}
where ${\cal S}^{X,Y,Z,U}_{4}(\rho,A,B,C,D)$ is defined as
\begin{eqnarray}
        &&\nonumber{\cal S}^{X,Y,Z,U}_{4}(\rho,A,B,C,D)
        \\
        &:=&\langle A_{0}B_{0}C_{0}D_{0}\rangle_{\rho}-\langle A_{1}B_{0}C_{0}D_{0}\rangle_{\rho}
        \nonumber \\ \nonumber&&-\langle A_{0}B_{1}C_{0}D_{0}\rangle_{\rho}-\langle A_{0}B_{0}C_{1}D_{0}\rangle_{\rho}
        \\ \nonumber&&-\langle A_{0}B_{0}C_{0}D_{1}\rangle_{\rho}-\langle A_{1}B_{1}C_{0}D_{0}\rangle_{\rho}
        \\ \nonumber&&-\langle A_{1}B_{0}C_{1}D_{0}\rangle_{\rho}-\langle A_{1}B_{0}C_{0}D_{1}\rangle_{\rho}
        \\ \nonumber&&-\langle A_{0}B_{1}C_{1}D_{0}\rangle_{\rho}-\langle A_{0}B_{1}C_{0}D_{1}\rangle_{\rho}
        \\ \nonumber&&-\langle A_{0}B_{0}C_{1}D_{1}\rangle_{\rho}+\langle A_{1}B_{1}C_{1}D_{0}\rangle_{\rho}
        \\ \nonumber&&+\langle A_{1}B_{1}C_{0}D_{1}\rangle_{\rho}+\langle A_{1}B_{0}C_{1}D_{1}\rangle_{\rho}
        \\ &&+\langle A_{0}B_{1}C_{1}D_{1}\rangle_{\rho}+\langle A_{1}B_{1}C_{1}D_{1}\rangle_{\rho}
\label{eq58}
\end{eqnarray}
Combining the inequality (\ref{eq60}) with Eq.(\ref{eq58}), we obtain that
\begin{eqnarray}
        & &{\cal S}^{X,Y,Z,U}_{4}(\rho,A,B,C,D)
        \nonumber\\
        &\leq&\langle A_{0}B_{0}\rangle(\langle C_{0}D_{0}\rangle-\langle C_{1}D_{0}\rangle-\langle C_{0}D_{1}\rangle-\langle C_{1}D_{1}\rangle)
      \nonumber\\
        &&+\langle A_{1}B_{1}\rangle(\langle C_{0}D_{0}\rangle-\langle C_{1}D_{0}\rangle-\langle C_{0}D_{1}\rangle-\langle C_{1}D_{1}\rangle)
         \nonumber\\
        &&+\langle A_{0}B_{1}\rangle(\langle C_{1}D_{1}\rangle-\langle C_{0}D_{0}\rangle-\langle C_{1}D_{0}\rangle-\langle C_{0}D_{1}\rangle)
         \nonumber\\
         &&+\langle A_{1}B_{0}\rangle(\langle C_{1}D_{1}\rangle-\langle C_{0}D_{0}\rangle-\langle C_{1}D_{0}\rangle-\langle C_{0}D_{1}\rangle)
         \nonumber\\
         &&+16 ce^{-\kappa\tau}|X|
         \nonumber\\
         &\leq & 2|\langle C_{0}D_{0}\rangle-\langle C_{1}D_{0}\rangle-\langle C_{0}D_{1}\rangle-\langle C_{1}D_{1}\rangle|
         \nonumber\\
         &&+2|\langle C_{1}D_{1}\rangle-\langle C_{0}D_{0}\rangle-\langle C_{1}D_{0}\rangle-\langle C_{0}D_{1}\rangle|
         \nonumber\\
         &&+16 ce^{-\kappa\tau}|X|
       \nonumber\\
         &\leq & 4\max\{|\langle C_{0}D_{0}\rangle|+|\langle C_{1}D_{1}\rangle|,\,|\langle C_{1}D_{0}\rangle|+|\langle C_{0}D_{1}\rangle|\}
         \nonumber\\
         &&+16 ce^{-\kappa\tau}|X|
         \nonumber\\
         &\leq & 8+16 ce^{-\kappa\tau}|X|
\label{eq61}
\end{eqnarray}
This implies that
\begin{eqnarray}
{\cal S}^{X,Y,Z,U}_{4}(\rho)\leq  8+16 ce^{-\kappa\tau}|X|
\label{eq65}
\end{eqnarray}

\textit{Example 8}. Consider the direct generalization of the Svetlichny tripartite inequality (\ref{eq3.20}). For an odd $n\geq4$ it similar to Theorem \ref{07}. For an even $n$, suppose that $\kappa$ is defined on exponentially clustering of correlations.  $E_{1},\cdots, E_{n}$ are supported in $X_{1},\cdots,X_{n}$, respectively. Combing Eqs.(\ref{eq3.20}) and (\ref{eq3.21}) with the inequalities (\ref{eq3.22}) and (\ref{eq69}), we get
\begin{eqnarray}
\nonumber |S_{n}^{\pm}| & \leq&2^{n-2}|CH_{1}|+ 2^{n-2}|CH_{2}| +(n-2)2^{n-1}ce^{-\kappa\tau}|X|
\\ \nonumber& \leq& 2^{n-2} \max\{ |CH_{1}+CH_{2}|, |CH_{1}-CH_{2}| \}
\\ \nonumber&& +(n-2)2^{n-1}ce^{-\kappa\tau}|X|
\\ & \leq& 2^{n-1} +(n-2)2^{n-1}ce^{-\kappa\tau}|X|
\label{eq3.30}
\end{eqnarray}

Theorem \ref{12} extends to the $n$-particle system by clustering $n$-particle systems into correlations of almost two bodies. It means that the quantum correlations for the gapped ground states will not significantly violate any $n$-partite Bell inequality (\ref{eq40}) when all the involved regions are far away from each other. This is different from recent result by clustering $N$-particle systems to $N$ uncorrelated parties \cite{Vieira}.

\section{Thermal states}

In this section, we will characterize thermal states at inverse temperature less than a fixed temperature $\beta^{*}$. The thermal state provides a valid description of the equilibrium state for many systems \cite{Kliesch}.

\subsection{Tripartite systems}

From Lemma \ref{02}, we prove the following result for tripartite systems.

\begin{theorem}
\label{05}
Let $\rho(\beta)$ be a thermal state acting on the lattice at inverse temperature $\beta$ less than a fixed constant $\beta^{*}$.  We get $\rho(\beta)$ is $\varepsilon$-local with  $\varepsilon=\eta c(\beta)e^{-\kappa(\beta)\tau}$.

\end{theorem}

\textbf{Proof}. It can be seen from Lemma \ref{02} that the thermal states satisfy exponentially clustering of correlations. From the inequality (\ref{eq3.7}) we obtain that
\begin{eqnarray}
\nonumber{\cal S}^{XYZ}(\rho,E_{k_{1}}^{(1)},E_{k_{2}}^{(2)},E_{k_{3}}^{(3)})
&\leq
& \hat{\cal S}^{XYZ}(\rho(\beta),E_{k_{1}}^{(1)},E_{k_{2}}^{(2)},E_{k_{3}}^{(3)})
\\
&& +\eta c(\beta)e^{-\kappa (\beta)\tau}
\label{eq4.1}
\end{eqnarray}

Suppose that $\hat{\cal S}^{X,Y,Z}(\rho(\beta),E_{k_{1}}^{(1)},E_{k_{2}}^{(2)},E_{k_{3}}^{(3)})$ is the expect of Bell operator based on any system with correlations of almost two bodies, which is defined in Eq.(\ref{eq3.2}). From the inequalities (\ref{eq3.1}) and (\ref{eq4.1}) we get that
\begin{eqnarray}
 \hat{\cal S}^{XYZ}(\rho,E_{k_{1}}^{(1)},E_{k_{2}}^{(2)},E_{k_{3}}^{(3)})
\leq \Delta_{loc} +\eta c(\beta)e^{-\kappa(\beta)\tau}
\label{eq4.2}
\end{eqnarray}
where $\Delta_{loc}$ is defined in the inequality (\ref{eq3.1}).

From the inequalities (\ref{eq3.1}), (\ref{eq4.1}) and (\ref{eq4.2}), it follows that
\begin{eqnarray}
\nonumber{\cal S}^{XYZ}(\rho,E_{k_{1}}^{(1)},E_{k_{2}}^{(2)},E_{k_{3}}^{(3)})
&\leq& \Delta_{loc} +2\eta c(\beta)e^{-\kappa(\beta)\tau}
\\
&\leq &\Delta_{loc}+\varepsilon
\end{eqnarray}
if $\tau$ is large enough, that is, the regions $X$, $Y$ and $Z$ are far away from each other. This means that there is almost no violation of tripartite bilocal inequalities. It completes the proof. $\square$

By using the additional properties of thermal states in specific Svetlichny scenario, we get a stronger result than Theorem \ref{05} as follows.

\textit{Example 9}. Consider the Svetlichny scenario. For any observables $A, B$ and $C$ supported respectively in $X$, $Y$ and $Z\subset\Omega$. It can be seen from Lemma \ref{02} that the thermal states satisfy exponentially clustering of correlations. From the inequalities (\ref{eq33}) and (\ref{eq4.1}) we obtain that
\begin{eqnarray}
\nonumber{\cal S}^{X,Y,Z}_{SI}(\rho(\beta),A,B,C)
&\leq
& \hat{\cal S}^{X,Y,Z}_{SI}(\rho(\beta),A,B,C)
\\
& &+8c(\beta)e^{-\kappa(\beta)\tau}
\label{eq81}
\end{eqnarray}
Define $B_{i}=b_{i}^{1}-b_{i}^{-1}$ and $C_{j}=c_{j}^{1}-c_{j}^{-1}$ for each $i,j\in\{0,1\}$, where $\{b^{-1}_{i},b^{1}_{i}\}$ and $\{c^{-1}_{j},c^{1}_{j}\}$ are positive-operator-value measurements associated with two measurement outcomes. Assume that $b^{1}_{i}$, $b^{-1}_{i}$, $c^{1}_{j}$, $c^{-1}_{j}\neq \mathbbm{1}$, and $0< {\rm tr}(b_{i}^{1}\otimes{}c_{j}^{1}\rho(\beta))$, ${\rm tr}(b_{i}^{-1}\otimes{}c_{j}^{-1}\rho(\beta))\leq 1$. It follows that
\begin{eqnarray}
\langle B_{i}C_{j}\rangle
 &=&{\rm tr}((b_{i}^{1}-b_{i}^{-1})\otimes(c_{j}^{1}-c_{j}^{-1})\rho(\beta))
\nonumber\\
&=&{\rm tr}(b_{i}^{1}\otimes(c_{j}^{1}-c_{j}^{-1})\rho(\beta))
\nonumber
\\
&&
-{\rm tr}(b_{i}^{-1}\otimes(c_{j}^{1}-c_{j}^{-1})\rho(\beta))
\label{eq82}
\end{eqnarray}
For the case of ${\rm tr}(b_{i}^{-1}\otimes(c_{j}^{1}-c_{j}^{-1})\rho(\beta))\geq0$, it follows that
\begin{eqnarray}
\langle B_{i}C_{j}\rangle
& \leq&   {\rm tr}(b_{i}^{1}\otimes(c_{j}^{1}-c_{j}^{-1})\rho(\beta))
\nonumber\\
 &<&{\rm tr}(b_{i}^{1}\otimes{}c_{j}^{1}\rho(\beta))
 \nonumber\\
 &<&1
\label{eq83}
\end{eqnarray}
Otherwise, we have ${\rm tr}(b_{i}^{-1}\otimes(c_{j}^{1}-c_{j}^{-1})\rho(\beta))<0$. This implies that
\begin{eqnarray}
\langle B_{i}C_{j}\rangle
 &<&{\rm tr}(b_{i}^{-1}\otimes(c_{j}^{1}-c_{j}^{-1})\rho(\beta))
 \nonumber\\
 &<&{\rm tr}(b_{i}^{-1}\otimes{}c_{j}^{-1}\rho(\beta))
  \nonumber\\
 &<&1
\label{eq84}
\end{eqnarray}
Similarly, we have
\begin{eqnarray}
-\langle B_{i}C_{j}\rangle<1
\label{eq86}
\end{eqnarray}
Assume that $\delta>0$. From the inequalities (\ref{eq81})-(\ref{eq86}) we get that
\begin{eqnarray}
{\cal S}^{X,Y,Z}_{SI}(\rho(\beta),A,B,C)\leq4-\delta+8c(\beta)e^{-\kappa(\beta)\tau}
\label{eq88}
\end{eqnarray}
If the regions $Y$ and $Z$ are far away from each other, there is $\tau^{*}=\frac{1}{\kappa}\ln(\frac{8c(\beta)}{\delta})$. Hence, we get
\begin{eqnarray}
{\cal S}^{X,Y,Z}_{SI}(\rho(\beta),A,B,C)  \leq4
\end{eqnarray}
for any $\tau\geq\tau^{*}$. It means that there is no violation of Svetlichny inequality \cite{Svetlichny} if all the involved regions are far away.

The genuinely multipartite states can be characterized according to rotational invariance \cite{Stasinska}. It is also useful for ground and thermal states defined on any lattice geometry \cite{Stasinska}. Especially, by clustering the systems into two independent parts in term of CHSH scenario \cite{Clauser}, the nonlocality cannot be observed in the experiment for two regions being far enough \cite{Vieira}. Here, we prove a thermal state $\rho$ being acted on the lattice at any inverse temperature $\beta<\beta^{*}$ is $\epsilon$-local \cite{Seevinck}. Combining with exponentially clustering of correlations in Lemma \ref{02}, from Example 9 there is a stronger result for thermal states than ground states, i.e., the biseparable bound is strictly preserved if the involved regions are far from each other.

\subsection{$n$-partite systems}

In this subsection, we consider the $n$-partite systems by using different clustering methods. We firstly introduce the following fact:

\textbf{Fact 3.} \emph{For any set of disjoint regions $X_{1},\cdot\cdot\cdot,X_{s}\subset\Omega$, suppose that a thermal state $\rho$ acting on ${\cal H}_\Omega$ shows exponentially clustering of correlations with $c$, $\tau>0$. We have
\begin{eqnarray}
  \nonumber&&|\langle E_{1}\cdot\cdot\cdot E_{s}\rangle_{\rho(\beta)}-\langle E_{1}\rangle_{\rho(\beta)}\cdot\cdot\cdot\langle E_{s-2}\rangle_{\rho(\beta)}\langle E_{s-1}E_{s}\rangle_{\rho(\beta)}|
  \\
  & \leq &(s-2) c(\beta)e^{-\kappa(\beta)\tau} \prod_{i=1}^s\|E_{i}\|
\end{eqnarray}
for any set of measurement operators $E_{1},\cdot\cdot\cdot,E_{s}$ supported in $X_{1},\cdot\cdot\cdot,X_{s}$ respectively, where $c$ and $\kappa$ are defined on exponentially clustering of correlations, $\tau=\min\{\tau_{ij}\}$, and $\tau_{ij}$ denotes the distance between the regions $X_{i}$ and $X_{j}$.}

The proof is similar to Fact 1 by using Lemma \ref{02}. From the inequality (\ref{eq40}) and Fact 3 we get the result for thermal states as follows.

\begin{theorem}\label{08}
Let $\rho(\beta)$ is a thermal state acting on the lattice at inverse temperature $\beta<\beta^{*}$. For every disjoint regions $X_{1},\cdots,X_{n}\subset\Omega$, there exist constants $c$, $\kappa>0$ such that $\rho(\beta)$ is $\varepsilon$-local with $\varepsilon= \gamma c(\beta)e^{-\kappa(\beta)\tau}$.

\end{theorem}

\textbf{Proof}. The proof is similar to Theorem \ref{07}. From Fact 3, we get that
\begin{eqnarray}
\nonumber&&\psi^{(i_{1}\cdots{}i_{s})}_{k_{1}\cdots{}k_{s}}\langle E^{(i_{1})}_{k_{1}}\cdots{} E^{(i_{s})}_{k_{s}}\rangle
\\ \nonumber&\leq&\psi^{(i_{1}\cdots{}i_{s})}_{k_{1}\cdots{}k_{s}}\langle E^{(i_{1})}_{k_{1}}\rangle\cdots\langle E^{(i_{s-2})}_{k_{s-2}}\rangle\langle E^{(i_{s-1})}_{k_{s-1}} E^{(i_{s})}_{k_{s}}\rangle
\\ && +(s-2)c(\beta)e^{-\kappa(\beta)\tau} |\psi^{(i_{1}\cdots{}i_{s})}_{k_{1}\cdots{}k_{s}}|
\label{eq89}
\end{eqnarray}
Combining with Eq.(\ref{eq42}), we get that
\begin{eqnarray}
\nonumber&&{\cal S}^{X_{1}\cdots{}X_{n}}(\rho(\beta),E^{(1)},\cdots,E^{(n)})
\\ &
\leq&\hat{\cal S}^{X_{1}\cdots{}X_{n}}(\rho(\beta),E^{(1)},\cdots,E^{(n)})
 +\gamma c(\beta)e^{-\kappa(\beta)\tau}
\label{eq90}
\end{eqnarray}
where $\hat{\cal S}^{X_{1}\cdots{}X_{n}}(\rho(\beta),E^{(1)},\cdots,E^{(n)})$ is defined in Eq.(\ref{eq52}). From the inequality (\ref{eq40}), it follows that
$\hat{\cal S}^{X_{1}\cdots{}X_{n}}(\rho(\beta),E^{(1)},\cdots,E^{(n)})\leq \Delta_{bi}$. So, the upper bound of ${\cal S}^{X_{1}\cdots{}X_{n}}(\rho(\beta),E^{(1)},\cdot\cdot\cdot,E^{(n)})$ in the inequality (\ref{eq90}) does not depend on the measurement operators $E^{(1)},\cdot\cdot\cdot,E^{(n)}$. This implies that
\begin{eqnarray}
{\cal S}^{X_{1}\cdots{}X_{n}}(\rho(\beta))&=&\sup_{E^{(1)},\cdots, E^{(n)}}{\cal S}^{X_{1}\cdots{}X_{n}}(\rho,E^{(1)},\cdots, E^{(n)})
\nonumber\\
&\leq&\Delta_{loc}+\gamma c(\beta)e^{-\kappa(\beta)\tau}
\label{eq91}
\end{eqnarray}
The proof is completed. $\square$

\textit{Example 10}. Considering the Bell-type inequalities for partial separability in $n$-particle systems in Eq.(\ref{eq3.20}). Let the set of operators and $E_{1},\cdots,E_{n}$ supported in $X_{1},\cdots,X_{n}$, and $\rho(\beta)$ is a thermal state acting on the lattice at inverse temperature $\beta<\beta^{*}$. Combining Eqs.(\ref{eq3.20})-(\ref{eq3.22}) with the inequality (\ref{eq89}), we can get
\begin{eqnarray}
\nonumber |S_{n}^{\pm}| & \leq&2^{n-2}|CH_{1}|+ 2^{n-2}|CH_{2}| +(n-2)2^{n-1}c(\beta)e^{-\kappa(\beta)\tau}
\\ \nonumber& \leq& 2^{n-2} \max\{ |CH_{1}+CH_{2}|, |CH_{1}-CH_{2}| \}
\\ \nonumber&& +(n-2)2^{n-1}c(\beta)e^{-\kappa(\beta)\tau}
\\ & \leq& 2^{n-1} +(n-2)2^{n-1}c(\beta)e^{-\kappa(\beta)\tau}
\label{eq4.23}
\end{eqnarray}

Theorem \ref{08} means that the bound in the inequality (\ref{eq40}) will not be significantly violated in experiments for thermal states if all the involved regions are far away from each other.

Similar to Lemma 4, we can show four-particle clustering method with an exponential decay for thermal states as follows:

\begin{lemma}\label{31}
Suppose a thermal state $\rho(\beta)$ at inverse temperature $\beta<\beta^{*}$. Let $A$, $B$, $C$ and $D\in{\cal B}({\cal H}_{\Omega})$ be the measurement operators being supported in disjoint regions $X$, $Y$, $Z$ and $U\subset\Omega$ in the lattice respectively. There exist $c,\kappa >0$ such that
\begin{eqnarray}
\nonumber|\langle ABCD\rangle_{\rho_{(\beta)}}-\langle AB\rangle_{\rho_{(\beta)}}\langle CD\rangle_{\rho_{(\beta)}}|&\leq\,& c(\beta)e^{-\kappa(\beta)\tau}\|A\|\,\|B\|
\\  &&\times{}\,\|C\|\,\|D\|
\label{eqlemma4.24}
\\ \nonumber|\langle ABCD\rangle_{\rho_{(\beta)}}-\langle AC\rangle_{\rho_{(\beta)}}\langle BD\rangle_{\rho_{(\beta)}}|&\leq\,& c(\beta)e^{-\kappa(\beta)\tau}\|A\|\,\|B\|
\\  &&\times{}\,\|C\|\,\|D\|
\label{eqlemma4.25}
\\ \nonumber|\langle ABCD\rangle_{\rho_{(\beta)}}-\langle AD\rangle_{\rho_{(\beta)}}\langle BC\rangle_{\rho_{(\beta)}}|&\leq\,& c(\beta)e^{-\kappa(\beta)\tau}\|A\|\,\|B\|
\\  &&\times{}\,\|C\|\,\|D\|
\label{eqlemma4.26}
\end{eqnarray}
where $\tau$ is the minimum distance between any two regions of $X$, $Y$, $Z$ and $U$.
\end{lemma}

\textbf{Proof}. Suppose that $\rho(\beta)$ is a thermal state at inverse temperature $\beta<\beta^{*}$. From Lemma \ref{02} we have
\begin{eqnarray}
|\langle ABC\rangle-\langle A\rangle\langle BC\rangle|  \leq c(\beta)e^{-\kappa(\beta)\tau'}
\label{eq4.21}
\end{eqnarray}
where $\tau'$ is the minimal distance between any two regions of $X$, $Y$ and $Z$. We take $X$ and $Y$ as one region, $Z$ and $U$ as respectively separate regions. Such we conclude by the definition of a state with exponential clustering of correlations:
\begin{eqnarray}
|\langle ABCD\rangle-\langle AB\rangle\langle CD\rangle|  \leq  c(\beta)e^{-\kappa(\beta)\tau}
\label{eq4.22}
\end{eqnarray}
where $\tau$ is the minimal distance between any two regions of $X$, $Y$, $Z$ and $U$. The same procedure may be easily adapted to obtain the inequalities (\ref{eqlemma4.25}) and (\ref{eqlemma4.26}). This completes the proof. $\square$

With this Lemma, we can prove the result for $n$-particle thermal states.

\textbf{Fact 4.} \emph{Assume that a thermal state $\rho(\beta)$ acting on ${\cal H}_\Omega$ shows exponentially clustering of correlations. For any set of disjoint regions $X_{1},\cdots,X_{s}\subset\Omega$, there exist constants $c,\kappa>0$ such that
\begin{eqnarray}
  \nonumber &&|\langle E_{1}\cdots E_{s}\rangle_{\rho(\beta)}-\langle E_{i_{1}}\cdots E_{i_{k}}\rangle_{\rho(\beta)}\langle E_{i_{k+1}}\cdots E_{i_{s}}\rangle_{\rho(\beta)}|
  \\
  \nonumber
  &\leq& c(\beta)e^{-\kappa(\beta)\tau} \prod_{i=1}^s\|E_{i}\|
\end{eqnarray}
for any set of operators $E_{1},\cdots,E_{s}$ acting respectively on $X_{1},\cdots,X_{s}$.}

\textbf{Proof}. About the bipartition of $\hat{X}_{1}=\{X_{i_{1}},\cdots,X_{i_{k}}\}$ and $\hat{X}_{2}=\{X_{i_{k+1}},\cdots,X_{i_{s}}\}$,
we have $d(\hat{X}_{1},\hat{X}_{2}) =\min \{ {d(X_{i},X_{j})} \}$ with $i\in \{ i_{1},\ldots,i_{k}\}$ and $j\in \{ i_{k+1},\ldots,i_{s} \}$. For any $i$ and $j$, we obtain $d(\hat{X}_{1},\hat{X}_{2}) \leq d(X_{i},X_{j}) \leq \tau$. It means that the minimal distance between $\cup_{i=i_{1}}^{i_{k}} X_{i}$ and $\cup_{i=i_{k+1}}^{i_{s}} X_{i}$ is no less than $\tau$. According to Lemma \ref{31}, we obtain that
\begin{eqnarray}
  \nonumber &&|\langle E_{1}\cdots E_{s}\rangle_{\rho(\beta)}-\langle E_{i_{i}}\cdots E_{i_{k}}\rangle_{\rho(\beta)}\langle E_{i_{k+1}}\cdots E_{i_{s}}\rangle_{\rho(\beta)}|
  \\
  &\leq& c(\beta)e^{-\kappa(\beta)\tau} \prod_{i=1}^s\|E_{i}\|
\label{eq66}
\end{eqnarray}
The proof is completed. $\square$

Based on Fact 4, we get the result for thermal states.

\begin{theorem}\label{13}
Let $\rho(\beta)$ be a thermal state acting on the lattice at inverse temperature $\beta<\beta^{*}$. There exist $c$, $\kappa>0$ such that $\rho(\beta)$ is $\varepsilon$-local with respect for any given disjoint regions $X_{1},\cdots,X_{n}\subset\Omega$, where $\varepsilon= \hat{\gamma} c(\beta)e^{-\kappa(\beta)\tau}$ and $\hat{\gamma}$ is given in Eq.(\ref{eq68}).

\end{theorem}

\textbf{Proof}. By Fact 4 and Eq.(\ref{eq42}) we get that
\begin{eqnarray}
\nonumber
 &&{\cal S}^{X_{1}\cdots{}X_{n}}(\rho(\beta),E^{(1)},\cdots,E^{(n)})
\\
 \nonumber
 &\leq&
 \tilde{\cal S}^{X_{1}\cdots{}X_{n}}(\rho(\beta),E^{(1)},\cdots,E^{(n)})
\\ \nonumber
 &&+c(\beta)e^{-\kappa(\beta)\tau}\sum^{n}_{s=1}\sum^{n}_{i_{1}
 \neq\cdots\neq i_{s}=1}
 \sum^{m_{i_{1}},\cdots,m_{i_{s}}}_{k_{1},\cdots,k_{s}=1}
 |\psi^{(i_{1}\cdots{}i_{s})}_{k_{1}\cdots{}k_{s}}|
 \\
 \label{eq93a}
\end{eqnarray}
where $\tilde{\cal S}^{X_{1}\cdots{}X_{n}}(\rho(\beta),E^{(1)},\cdots,E^{(n)})$ is defined in Eq.(\ref{eq71}). From the inequality (\ref{eq40}), it follows that
$\tilde{\cal S}^{X_{1}\cdots{}X_{n}}(\rho(\beta),E^{(1)},\cdots,E^{(n)})\leq \Delta_{bi}$. From the inequality (\ref{eq93a}) it follows that
\begin{eqnarray}
\nonumber
 &&{\cal S}^{X_{1}\cdots{}X_{n}}(\rho(\beta),E^{(1)},\cdots,E^{(n)})
 \\
&\leq&\Delta_{loc}+\hat{\gamma}c(\beta)e^{-\kappa(\beta)\tau}
\label{eq93}
\end{eqnarray} 
where $\Delta_{loc}$ is defined in the inequality (\ref{eq40}). So, the upper bound of ${\cal S}^{X_{1}\cdots{}X_{n}}(\rho(\beta),E^{(1)},\cdot\cdot\cdot,E^{(n)})$ in the inequality (\ref{eq93}) does not depend on the measurement operators $E^{(1)},\cdot\cdot\cdot,E^{(n)}$. This implies that
\begin{eqnarray}
{\cal S}^{X_{1}\cdots{}X_{n}}(\rho(\beta))
&=&
\sup_{E^{(1)},\cdots, E^{(n)}}{\cal S}^{X_{1}\cdots{}X_{n}}(\rho,E^{(1)},\cdots, E^{(n)})
\nonumber\\
&\leq&
\Delta_{loc}+\hat{\gamma} c(\beta)e^{-\kappa(\beta)\tau}
\label{eq94}
\end{eqnarray}
It completes the proof. $\square$

\textit{Example 11}. From Lemma \ref{31}, we know that $\rho(\beta)$ is a thermal state satisfying exponentially clustering of correlations. Hence, there is $\delta'>0$ such that
\begin{eqnarray}
        \nonumber{\cal S}^{XYZU}_{4}(\rho(\beta),A,B,C,D)&\leq
        & \hat{\cal S}^{XYZU}_{4}(\rho(\beta),A,B,C,D)
        \\
        \nonumber
        && +16c(\beta)e^{-\kappa(\beta)\tau}
        \\
        \nonumber
        &\leq
        &\,8-\delta'+16c(\beta)e^{-\kappa(\beta)\tau}
        \\
\label{eq92}
\end{eqnarray}
from Theorem \ref{13} and Example 9, where $\tau^{*}=\frac{1}{\kappa}\ln(\frac{16c(\beta)}{\delta})$. We get ${\cal S}^{XYZU}_{4}(\rho(\beta),A,B,C,D)\leq 8$ for $\tau\geq\tau^{*}$.

\textit{Example 12}. For the $n$-particle Bell-type inequalities under the assumption of partial separability as in Eq.(\ref{eq3.20}). Suppose that $\rho(\beta)$ is a thermal state acting on the lattice at inverse temperature $\beta<\beta^{*}$.  For any observables $E_{1},\cdots,E_{n}$ supported in disjoint regions $X_{1}, \cdots, X_{n}$, from Eqs.(\ref{eq3.20})-(\ref{eq3.22}) and the inequality (\ref{eq93a}), we can get
\begin{eqnarray}
\nonumber |S_{n}^{\pm}| & \leq&2^{n-2}|CH_{1}|+ 2^{n-2}|CH_{2}| +(n-2)2^{n-1}c(\beta)e^{-\kappa(\beta)\tau}
\\ \nonumber& \leq& 2^{n-2} \max\{ |CH_{1}+CH_{2}|, |CH_{1}-CH_{2}| \}
\\ \nonumber&& +(n-2)2^{n-1}c(\beta)e^{-\kappa(\beta)\tau}
\\ & \leq& 2^{n-1} +(n-2)2^{n-1}c(\beta)e^{-\kappa(\beta)\tau}
\label{eq4.24}
\end{eqnarray}

From Theorems \ref{08} and \ref{13}, the thermal states at inverse temperature less than a fixed temperature $\beta^{\ast}$ do not significant violate any Bell inequality (\ref{eq40}) when all the involved regions are far away from each other. This means that these thermal states show almost no obvious genuinely multipartite nonlocality by using two kinds of exponentially clustering correlations. These are going beyond recent results with different clustering theorems  \cite{Vieira}.

\section{Product states as initial states}

In previous sections, we have proved that there is an exponential decay of correlations with the distance. This restricts the propagation of the association in the lattice when the initial states are gapped ground states or thermal states. In this section, the main result is related to product states as the initial states.

\subsection{Tripartite systems}

In this subsection, we consider the tripartite system. Suppose that the initial state of the system is a product state, that is, $\rho(0)=\otimes_{x\epsilon\Omega}\rho_{x}$.

\begin{theorem}
\label{06}
There are $c, \kappa, v>0$ such that for given three disjoint regions $X$, $Y$ and $Z\subset\Omega$, $\rho(t)$ is $\varepsilon$-local with $\varepsilon=\eta ce^{-\kappa\tau}(e^{\kappa v t}-1)|X|\,|Y|\,|Z|$.

\end{theorem}

\textbf{Proof}. Similar to the proof of Theorem \ref{04}, from Lemma \ref{03} we obtain
\begin{eqnarray}
\nonumber&&\psi_{k_{1}k_{2}k_{3}}\langle E_{k_{1}}^{(1)},E_{k_{2}}^{(2)},E_{k_{3}}^{(3)}\rangle
\\ \nonumber&\leq& \psi_{k_{1}k_{2}k_{3}} \langle E^{(1)}_{k_{1}}\rangle\langle E^{(2)}_{k_{2}} E^{(3)}_{k_{3}}\rangle
\\ &&+ce^{-\kappa\tau}(e^{\kappa v t}-1) |\psi_{k_{1}k_{2}k_{3}}| \,|X|\,|Y|\,|Z|
\label{eq5.1}
\end{eqnarray}
Combined with Eq.(\ref{eq3.2}), we have
\begin{eqnarray}
\nonumber{\cal S}^{XYZ}(\rho,E_{k_{1}}^{(1)},E_{k_{2}}^{(2)},E_{k_{3}}^{(3)})&\leq
 & \hat{\cal S}^{XYZ}(\rho,E_{k_{1}}^{(1)},E_{k_{2}}^{(2)},E_{k_{3}}^{(3)})
\\ \nonumber
&&+\eta ce^{-\kappa\tau}(e^{\kappa v t}-1) |X|\,|Y|\,|Z|
\\ \nonumber
&\leq & \Delta_{loc}+ \eta ce^{-\kappa\tau}(e^{\kappa v t}-1)
\\
&& \times{}|X|\,|Y|\,|Z|
\label{eq102a}
\end{eqnarray}
where $\hat{\cal S}^{XYZ}(\rho,E_{k_{1}}^{(1)},E_{k_{2}}^{(2)},E_{k_{3}}^{(3)})$ as the expect of Bell operators with correlations of almost two bodies defined in Eq.(\ref{eq3.6}). From the inequality (\ref{eq3.1}), it follows the inequality (\ref{eq102a}). This completes the proof. $\square$

\textit{Example 13}. For given three disjoint regions $X$, $Y$ and $Z\subset\Omega$, $\rho(t)$ is $\varepsilon$-local for Svetlichny inequality (\ref{eq31}). From Lemma \ref{03}, we get that
\begin{eqnarray}
\nonumber
\langle A_{i}B_{j}C_{k}\rangle\leq\langle A_{i}\rangle\langle B_{j}C_{k}\rangle+ ce^{-\kappa\tau}(e^{\kappa v t}-1)|X|\,|Y|\,|Z|
\\
\label{eq102b}
\end{eqnarray}
Combined with Eq.(\ref{eq33}), we have
\begin{eqnarray}
\nonumber{\cal S}^{XYZ}_{SI}(\rho(t),A,B,C)&\leq
 & \hat{\cal S}^{XYZ}_{SI}(\rho(t),A,B,C)
\\
\nonumber
&& +8 ce^{-\kappa\tau}(e^{\kappa v t}-1) |X|\,|Y|\,|Z|
\\
\nonumber
&\leq
& 4+8 ce^{-\kappa\tau}(e^{\kappa v t}-1) |X|\,|Y|\,|Z|
\\
\label{eq102c}
\end{eqnarray}

Here, we have introduced the Lieb-Robinson velocity $v$, which represents the maximum effective velocity at information travels on the lattice. Theorem \ref{06} implies that there does not exist genuinely tripartite nonlocality for the system in the time of $\frac{\tau}{v}$. This is different from the nonlocality detection for bipartition systems \cite{Vieira} by using the CHSH inequality \cite{Clauser}.

\subsection{$n$-partite systems}

For any disjoint regions $X_{1},\cdots,X_{s}\subset\Omega$, suppose that the initial state of the system is a product state $(\rho(0)=\otimes_{x\epsilon\Omega}\rho_{x})$. We introduce the following fact.

\textbf{Fact 5}. \emph{There are constants $c,\kappa,v>0$ such that for any set of operators $E_{1},\cdots,E_{s}$ supported respectively in $X_{1},\cdot\cdot\cdot,X_{s}$ we have
\begin{eqnarray}
  \nonumber&&|\langle E_{1}\cdots{}E_{s}\rangle_{\rho(t)}-\langle E_{1}\rangle_{\rho(t)}\cdots\langle E_{s-2}\rangle_{\rho(t)}\langle E_{s-1}E_{s}\rangle_{\rho(t)}|
  \\ &\leq& \alpha ce^{-\kappa\tau}(e^{\kappa v t}-1)  \prod_{i=1}^s\|E_{i}\|
\end{eqnarray}
where $\alpha=\Sigma^{s-2}_{i=1}\prod_{j=i}^s|X_{j}|$, $\tau=\min\{\tau_{ij}\}$, and $\tau_{ij}$ denotes the distance between the regions $X_{i}$ and $X_{j}$.}

The proof of Fact 5 is shown in Appendix B. With this fact we get the following result for general systems.

\begin{theorem}
\label{20} There exist constants $c,\kappa,v>0$ such that for every disjoint regions $X_{1},\cdots,X_{n}\subset\Omega$, $\rho(t)$ is $\varepsilon$-local with $\varepsilon= \mu ce^{-\kappa\tau}(e^{\kappa v t}-1)$ and $\mu=\sum^{n}_{s=1}\sum^{n}_{i_{1}\neq\cdots\neq i_{s}=1}\sum_{k_{1},\cdots,k_{s}}\alpha
|\psi^{(i_{1}\cdots{}i_{s})}_{k_{1}\cdots{}k_{s}}|$.

\end{theorem}

\textbf{Proof}. The proof is similar to Theorem \ref{07}. By Fact 5 we get
\begin{eqnarray}
\nonumber\psi^{(i_{1}\cdots{}i_{s})}_{k_{1}\cdots{}k_{s}}\langle E^{(i_{1})}_{k_{1}}\cdots E^{(i_{s})}_{k_{s}}\rangle &\leq&  \, \alpha ce^{-\kappa\tau}(e^{\kappa v t}-1)  |\psi^{(i_{1}\cdots{}i_{s})}_{k_{1}\cdots{}k_{s}}
|
\\
\nonumber&&+\psi^{(i_{1}\cdots{}i_{s})}_{k_{1}\cdots{}k_{s}}\langle E^{(i_{s-1})}_{k_{s-1}} E^{(i_{s})}_{k_{s}}\rangle
\\
&&\times \langle E^{(i_{1})}_{k_{1}}\rangle\cdots\langle E^{(i_{s-2})}_{k_{s-2}}\rangle
\label{eq5.2}
\end{eqnarray}
From Eq.(\ref{eq42}) we obtain that
\begin{eqnarray}
&&\nonumber{\cal S}^{X_{1}\cdots{}X_{n}}(\rho(t),E^{(1)},\cdots,E^{(n)})
\\
\nonumber
&\leq
&\hat{\cal S}^{X_{1}\cdots{}X_{n}}(\rho(t),E^{(1)},\cdots{},E^{(n)})
+\mu ce^{-\kappa\tau}(e^{\kappa v t}-1)
\\
\label{eq5.3}
\end{eqnarray}
where $\tilde{\cal S}^{X_{1}\cdots{}X_{n}}(\rho(\beta),E^{(1)},\cdots,E^{(n)})$ is defined in Eq.(\ref{eq52}). From the inequality (\ref{eq40}), it follows that
$\hat{\cal S}^{X_{1}\cdots{}X_{n}}(\rho(t),E^{(1)},\cdots{},E^{(n)})\leq \Delta_{bi}$. So, from the inequality (\ref{eq5.3}), we get
\begin{eqnarray}
&&\nonumber{\cal S}^{X_{1}\cdots{}X_{n}}(\rho(t),E^{(1)},\cdots,E^{(n)})
\\
&\leq
&\Delta_{loc}+\mu ce^{-\kappa\tau}(e^{\kappa v t}-1)
\label{eq5.3a}
\end{eqnarray}
which means that the upper bound of $\hat{\cal S}^{X_{1},\cdot\cdot\cdot,X_{n}}(\rho(t),E^{(1)},\cdot\cdot\cdot,E^{(n)})$ does not depend on the measurement operators $E^{(1)},\cdots{},E^{(n)}$. This implies that
\begin{eqnarray}
{\cal S}^{X_{1}\cdots{}X_{n}}(\rho(\beta))
&=&
\sup_{E^{(1)},\cdots, E^{(n)}}{\cal S}^{X_{1}\cdots{}X_{n}}(\rho,E^{(1)},\cdots, E^{(n)})
\nonumber\\
&\leq&\Delta_{loc}+\mu ce^{-\kappa\tau}(e^{\kappa v t}-1)
\label{eq5.4}
\end{eqnarray}
where $\Delta_{loc}$ is defined in the inequality (\ref{eq40}). It completes the proof. $\hfill\square$

\textit{Example 14}. Consider an $n$-particle Bell-type inequality under the assumption of partial separability as in Eq.(\ref{eq3.20}). Let the initial state of a lattice be $\rho(0)=\otimes_{x\epsilon\Omega}\rho_{x}$. For disjoint regions $X_{1},\cdots,X_{n}\subset\Omega$, from Eqs.(\ref{eq3.21})-(\ref{eq3.22}) and the inequality (\ref{eq5.2}) we get that
\begin{eqnarray}
\nonumber |S_{n}^{\pm}| & \leq &2^{n-2}|CH_{1}|+ 2^{n-2}|CH_{2}|
\\ \nonumber && +(n-2)2^{n} \alpha ce^{-\kappa\tau}(e^{\kappa v t}-1)
\\ \nonumber& \leq& 2^{n-2} \max\{ |CH_{1}+CH_{2}|, |CH_{1}-CH_{2}| \}
\\ \nonumber&& +(n-2)2^{n} \alpha ce^{-\kappa\tau}(e^{\kappa v t}-1)
\\ & \leq& 2^{n-1} +(n-2)2^{n} \alpha ce^{-\kappa\tau}(e^{\kappa v t}-1)
\label{eq5.21}
\end{eqnarray}

Note that $\alpha$ can be very large. However, it should be a constant with an upper bound in our hypothesis. The ploytope of genuinely multipartite correlations can be featured via linear inequalities (\ref{eq40}). This means that there is no obvious genuinely multipartite nonlocality in the time of $\frac{\tau}{v}$ when the involved regions are far apart.

For the four-particle scenario, we give two clustering methods to explore the genuinely multipartite nonlocality. For any four disjoint regions $X$, $Y$, $Z$ and $U\subset\Omega$, let $A$, $B$, $C$ and $D\in{\cal B}({\cal H}_{\Omega})$ be measurement operators acting on $X$, $Y$, $Z$ and $U$ respectively. Similar to Lemma 1, we get the bounds for different partitions.

\begin{lemma}\label{32}
Assume that the initial state of the lattice is $\rho(0)=\otimes_{x\in\Omega}\rho_{x}$. There exist constants $c,\tau>0$ such that 
\begin{eqnarray}
\nonumber&& |\langle ABCD\rangle_{\rho_{(t)}}-\langle AB\rangle_{\rho_{(t)}}\langle CD\rangle_{\rho_{(t)}}|
\\ &\leq& ce^{-\kappa\tau}(e^{\kappa v t}-1) \|A\|\,\|B\|\,\|C\|\,\|D\|\,|X|\,|Y|\,|Z|\,|U|
\nonumber\\
\label{eqlemma6.1}
\\ \nonumber&& |\langle ABCD\rangle_{\rho_{(t)}}-\langle AC\rangle_{\rho_{(t)}}\langle BD\rangle_{\rho_{(t)}}|
\\ &\leq& ce^{-\kappa\tau}(e^{\kappa v t}-1) \|A\|\,\|B\|\,\|C\|\,\|D\|\,|X|\,|Y|\,|Z|\,|U|
\nonumber\\
\label{eqlemma6.2}
\\ \nonumber&& |\langle ABCD\rangle_{\rho_{(t)}}-\langle AD\rangle_{\rho_{(t)}}\langle BC\rangle_{\rho_{(t)}}|
\\ &\leq& ce^{-\kappa\tau}(e^{\kappa v t}-1) \|A\|\,\|B\|\,\|C\|\,\|D\|\,|X|\,|Y|\,|Z|\,|U|
\nonumber\\
\label{eqlemma6.3}
\end{eqnarray}
where $\tau$ is the minimum distance between any two regions of $X$, $Y$, $Z$ and $U$.

\end{lemma}

\textbf{Proof}. Assume that the initial state of the lattice is $\rho(0)=\otimes_{x\in\Omega}\rho_{x}$. From Lemma \ref{03} we get
\begin{eqnarray}
 |\langle ABC\rangle-\langle A\rangle\langle BC\rangle|  \leq ce^{-\kappa\tau'}(e^{\kappa v t}-1)  |X|\,|Y|\,|Z|
\label{eq5.22}
\end{eqnarray}
where $\tau'$ is the minimal distance between any two regions of $X$, $Y$ and $Z$. Consider $X$ and $Y$ as one region, $Z$ and $U$ as respectively separate regions. We conclude by the definition of a state with exponential clustering of correlations as
\begin{eqnarray}
 |\langle ABCD\rangle-\langle AB\rangle\langle CD\rangle|  &\leq   & ce^{-\kappa\tau}(e^{\kappa v t}-1)  |X|
  \nonumber\\
  &&\times \,|Y|\,|Z|\,|U|
\label{eq5.23}
\end{eqnarray}
where $|X|=\max\{|X|,|Y|,|Z|,|U|\}$ and $\tau$ is the minimal distance between any two regions of $X$, $Y$, $Z$ and $U$. Similarly, we can prove the inequalities (\ref{eqlemma6.2}) and (\ref{eqlemma6.3}). This completes the proof. $\square$

Lemma \ref{32} is useful for proving the $\varepsilon$-bilocality of general states, where the initial state of the lattice is a product state $\rho(0)=\otimes_{x\epsilon\Omega}\rho_{x}$.

\textbf{Fact 6.} \textit{For any set of disjoint regions $X_{1},\cdot\cdot\cdot,X_{s}\subset\Omega$, there are constants $c,\kappa,v>0$ such that
\begin{eqnarray}
  \nonumber &&|\langle E_{1}\cdot\cdot\cdot E_{s}\rangle_{\rho(t)}-\langle E_{i_{1}}\cdot\cdot\cdot E_{i_{k}}\rangle_{\rho(t)}\langle E_{i_{k+1}}\cdot\cdot\cdot E_{i_{s}}\rangle_{\rho(t)}|
  \\
  &\leq& \xi ce^{-\kappa\tau}(e^{\kappa v t}-1) \prod_{i=1}^s\|E_{i}\|
  \label{Fact8}
\end{eqnarray}
where $|X|=\max\{|X_{1}|,\cdots,|X_{s}|\}$ and $\xi=\prod_{i=1}^{s}|X_{i}|$,  $\tau=\min\{\tau_{ij}\}$, and $\tau_{ij}$ is the distance between the regions $X_{i}$ and $X_{j}$.}

\textbf{Proof}. For the bipartition of $\hat{X}_{1}=\{X_{i_{1}},\cdots,X_{i_{k}}\}$ and $\hat{X}_{2}=\{X_{i_{k+1}},\cdots,X_{i_{s}}\}$,
we have $d(\hat{X}_{1},\hat{X}_{2}) =\min \{ {d(X_{i},X_{j})} \}$ with $i\in \{ i_{1},\cdots,i_{k} \}$ and $j\in \{ i_{k+1},\cdots,i_{s} \}$. For any $i$ and $j$, we obtain $d(\hat{X}_{1},\hat{X}_{2}) \leq d(X_{i},X_{j}) \leq \tau$. It means that the minimal distance between $\cup_{i=i_{1}}^{i_{k}} X_{i}$ and  $\cup_{i=i_{k+1}}^{i_{s}} X_{i}$ is no less than $\tau$. According to Lemma \ref{32}, we have
\begin{eqnarray}
  \nonumber &&|\langle E_{1}\cdot\cdot\cdot E_{s}\rangle_{\rho(t)}-\langle E_{i_{1}}\cdot\cdot\cdot E_{i_{k}}\rangle_{\rho(t)}\langle E_{i_{k+1}}\cdot\cdot\cdot E_{i_{s}}\rangle_{\rho(t)}|
  \\
  &\leq& \xi ce^{-\kappa\tau}(e^{\kappa v t}-1) \prod_{i=1}^s\|E_{i}\|
  \label{fact6}
\end{eqnarray}
This completes the proof. $\square$

\begin{theorem}\label{21}
There are constants $c,\kappa,v>0$ such that $\rho(t)$ is $\varepsilon$-local for any given disjoint regions $X_{1},\cdots,X_{n}\subset\Omega$, where $\varepsilon= \hat{\mu} ce^{-\kappa\tau}(e^{\kappa v t}-1)$,  $\hat{\mu}=\sum^{n}_{s=1}\sum^{n}_{i_{1}\neq\cdots\neq i_{s}=1}\sum_{k_{1},\cdots,k_{s}=1}
\xi |\psi^{(i_{1}\cdots{}i_{s})}_{k_{1}\cdots{}k_{s}}|$, and $\xi=\prod_{i=1}^{s}|X_{i}|$.

\end{theorem}

\textbf{Proof}. From Fact 6 and the inequality (\ref{eq42}), we get that
\begin{eqnarray}
&&{\cal S}^{X_{1}\cdots{}X_{n}}(\rho(t),E^{(1)},\cdots,E^{(n)})
\nonumber
\\
&\leq&\hat{\cal S}^{X_{1}\cdots{}X_{n}}(\rho(t),E^{(1)},\cdots,E^{(n)})
 \nonumber \\
 &&+ ce^{-\kappa\tau}(e^{\kappa v t}-1)\sum^{n}_{s=1}
 \sum^{n}_{i_{1}\neq\cdots\neq i_{s}=1}\sum_{k_{1},\cdots,k_{s}}\xi
 |\psi^{(i_1\cdots{}i_s)}_{k_{1}\cdots{}k_{s}}|
 \nonumber \\
 \label{eq107a}
\end{eqnarray}
where $\hat{\cal S}^{X_{1}\cdots{}X_{n}}(\rho(\beta),E^{(1)},\cdots,E^{(n)})$ is defined in Eq.(\ref{eq71}). From the inequality (\ref{eq40}), it follows that
$\hat{\cal S}^{X_{1}\cdots{}X_{n}}(\rho(t),E^{(1)},\cdots{},E^{(n)})\leq \Delta_{bi}$. So, from the inequality (\ref{eq107a}) we get
\begin{eqnarray}
&&{\cal S}^{X_{1}\cdots{}X_{n}}(\rho(t),E^{(1)},\cdots,E^{(n)})
\nonumber
\\
 &\leq&
 \Delta_{loc}+\hat{\mu} ce^{-\kappa\tau}(e^{\kappa v t}-1)
\label{eq107}
\end{eqnarray}
where $\Delta_{loc}$ is defined in the inequality (\ref{eq40}). According to Eq.(\ref{eq107}), the upper bound of ${\cal S}^{X_{1}\cdots{}X_{n}}(\rho(t))$ is independent of $E^{(1)},\cdots,E^{(n)}$. It follows that
\begin{eqnarray}
{\cal S}^{X_{1}\cdots{}X_{n}}(\rho(t))&=&\sup_{E^{(1)},\cdots, E^{(n)}}{\cal S}^{X_{1}\cdots{}X_{n}}(\rho,E^{(1)},\cdots, E^{(n)})
\nonumber\\
&\leq&\Delta_{loc}+\hat{\mu} ce^{-\kappa\tau}(e^{\kappa v t}-1)
\label{eq108}
\end{eqnarray}
which completes the proof. $\square$

\textit{Example 15}. Suppose the initial state of the system is a product state $(\rho(0)=\otimes_{x\epsilon\Omega}\rho_{x})$. According to Lemma \ref{32} we have
\begin{eqnarray}
\langle A_{i}B_{j}C_{k}D_{l}\rangle
&\leq& ce^{-\kappa\tau}(e^{\kappa v t-1}) |X|\,|Y|\,|Z|\,|U|
\nonumber\\
&& +\langle A_{i}B_{j}\rangle\langle C_{k}D_{l}\rangle
\label{eq5.24}
\end{eqnarray}
From Eq.$(\ref{eq58})$ we get that
\begin{eqnarray}
&&{\cal S}^{X,Y,Z,U}_{4}(\rho(t),A,B,C,D)
\nonumber\\
&\leq &\hat{\cal S}^{X,Y,Z,U}_{4}(\rho(t),A,B,C,D)
\nonumber\\
& &
+16 ce^{-\kappa\tau}(e^{\kappa v t-1}) |X|\,|Y|\,|Z|\,|U|
\nonumber\\
 &\leq&
 8+16 ce^{-\kappa\tau}(e^{\kappa v t-1}) |X|\,|Y|\,|Z|\,|U|
\label{eq5.25}
\end{eqnarray}

\textit{Example 16}. For the direct generalization of the Svetlichny tripartite inequalities in Eq.(\ref{eq3.20}). Assume that $\rho(0)=\otimes_{x\epsilon\Omega}\rho_{x}$ is the initial state of a lattice. Combing Eqs.(\ref{eq3.21})-(\ref{eq3.22}) and the inequality (\ref{eq107}) to get that
\begin{eqnarray}
\nonumber |S_{n}^{\pm}| &\leq& 2^{n-2}|CH_{1}|+ 2^{n-2}|CH_{2}|
\\ \nonumber && +(n-2)2^{n-2} \alpha ce^{-\kappa\tau}(e^{\kappa v t}-1)
\\ \nonumber& \leq& 2^{n-2} \max\{ |CH_{1}+CH_{2}|, |CH_{1}-CH_{2}| \}
\\ \nonumber&& +(n-2)2^{n-2} \alpha ce^{-\kappa\tau}(e^{\kappa v t}-1)
\\ & \leq &2^{n-1} +(n-2)2^{n-2} \alpha ce^{-\kappa\tau}(e^{\kappa v t}-1)
\label{eq5.26}
\end{eqnarray}

From Theorems \ref{20} and \ref{21}, since all the parties are far enough, ${\cal S}^{X_{1}\cdots{}X_{n}}(\rho(t))$ has an upper bound that is very close to the bound of the biseparable quantum correlations. This implies that one cannot verify the genuinely  multipartite nonlocality in a time of the order $\frac{\tau}{v}$.

\section{Conclusion}

In this paper, we investigated the genuinely multipartite nonlocality of many-body systems defined in lattice. The genuinely multipartite nonlocal correlations can be regarded as a convex combination of finite extremum points. The linear facet inequalities are used to characterise this polytope. We proved that there is almost no genuinely multipartite entanglement in the relevant classes of quantum states by using clustering theorems. In the spin lattices, we firstly explored the tripartite system. When three parties in the system acted on remote regions of the lattice, the ground state of the gap Hamiltonian cannot significant violate any tripartite Bell-bilocal inequalities. Moreover, for the thermal state a more restrictive conclusion in the specific scenario (as in Example 9) shows that there is a minimum distance between regions that rules out of any genuinely nonlocal correlations. For product states as initial states, the propagation relation of genuinely multipartite nonlocality is obtained over time. These results are further extended to $n$-partite systems by using different clustering theorems. These results can be used as a necessary condition to determine how to prepare many-body entanglement in the spin system. The present results are interesting in quantum many-body systems, quantum entanglement and quantum information processing.

\section*{Acknowledgements}

This work was supported by the national natural Science Foundation of China (no.61772437), and Fundamental Research Funds for the Central Universities (no.2682014CX095).

\appendix

\section{Proof of Fact 1}

Since $\rho$ is a state of exponentially clustering of correlations, from Lemma \ref{01} we have
\begin{eqnarray}
 |\langle E_{i}E_{j}E_{k}\rangle-\langle E_{i}\rangle\langle E_{j} E_{k}\rangle|
&\leq& ce^{-\kappa\tau'}|X'|
\nonumber\\
&\leq& ce^{-\kappa\tau}|X|
\label{eq43}
\end{eqnarray}
where $i,j,k\in\{1,\cdot\cdot\cdot,n\}$, $|X'|=\max\{|E_{i}|,|E_{j}|,|E_{k}|\}$ and $\tau'=\min\{\tau_{ij},\tau_{ik},\tau_{jk}\}$. With this result, we show Fact 1 by induction.

The case for $k=3$ is followed from Eq.(\ref{eq43}). For $k=m<n$, assume that we have
\begin{eqnarray}
 &&\nonumber|\langle E_{i_{1}}\cdots E_{i_{m}}\rangle-\langle E_{i_{1}}\rangle\cdots\langle E_{i_{m-2}}\rangle\langle E_{i_{m-1}}E_{i_{m}}\rangle|
 \nonumber
 \\
 &\leq& (m-2)ce^{-\kappa\tau} |X|
\label{eq44}
\end{eqnarray}
Now, consider the case of $k=m+1$. In fact, we get that
\begin{eqnarray}
 \langle E_{i_{1}}\cdots E_{i_{m+1}}\rangle
 &\leq&\langle E_{i_{1}}\rangle\cdots\langle E_{i_{m-2}}\rangle\langle E_{i_{m-1}}E_{i_{m}}E_{i_{m+1}}\rangle
  \nonumber\\
 &&+(m-2)ce^{-\kappa\tau} |X|
\nonumber \\
 &\leq & \langle E_{i_{1}}\rangle\cdots\langle E_{i_{m-2}}\rangle(\langle E_{i_{m-1}}\rangle\langle E_{i_{m}}E_{i_{m+1}}\rangle
  \nonumber\\
 && +ce^{-\kappa\tau}) |X|
+(m-2)ce^{-\kappa\tau} |X|
 \nonumber\\
 &\leq &\langle E_{i_{1}}\rangle\cdots\langle E_{i_{m-1}}\rangle\langle E_{i_{m}}E_{i_{m+1}}\rangle
  \nonumber\\
 &&+(m-1)ce^{-\kappa\tau} |X|
\label{eq45}
\end{eqnarray}
and
\begin{eqnarray}
\langle E_{i_{1}}\cdots E_{i_{m+1}}\rangle
 &\geq & \langle E_{i_{1}}\rangle\cdots\langle E_{i_{m-2}}\rangle\langle E_{i_{m-1}}E_{i_{m}}E_{i_{m+1}}\rangle
  \nonumber\\
 &&-(m-2)ce^{-\kappa\tau} |X|
 \nonumber\\
 &\geq &\langle E_{i_{1}}\rangle\cdots\langle E_{i_{m-2}}\rangle(\langle E_{i_{m-1}}\rangle\langle E_{i_{m}}E_{i_{m+1}}\rangle
  \nonumber\\
 &&-ce^{-\kappa\tau}) |X|
-(m-2)ce^{-\kappa\tau} |X|
 \nonumber\\
 &\geq &\langle E_{i_{1}}\rangle\cdots\langle E_{i_{m-1}}\rangle\langle E_{i_{m}}E_{i_{m+1}}\rangle
  \nonumber\\
 &&-(m-1)ce^{-\kappa\tau} |X|
\label{eq46}
\end{eqnarray}
Combining the inequalities (\ref{eq45}) and (\ref{eq46}), we get that
\begin{eqnarray}
(1-m)|X|ce^{-\kappa\tau}
 &\leq &\langle E_{i_{1}}\cdot\cdot\cdot E_{i_{m+1}}\rangle
  \nonumber\\
 &&
 -\langle E_{i_{1}}\rangle\cdot\cdot\cdot\langle E_{i_{m-1}}\rangle\langle E_{i_{m}}E_{i_{m+1}}\rangle
 \nonumber\\
 &\leq & (m-1)ce^{-\kappa\tau} |X|
\label{eq47}
\end{eqnarray}
Therefore, it implies that
\begin{eqnarray}
&&|\langle E_{i_{1}}\cdot\cdot\cdot E_{i_{s}}\rangle-\langle E_{i_{1}}\rangle\cdot\cdot\cdot\langle E_{i_{s-2}}\rangle\langle E_{i_{s-1}}E_{i_{s}}\rangle|
\nonumber
\\
&\leq & (s-2)ce^{-\kappa\tau} |X|
\label{eq48}
\end{eqnarray}

\section{The proof of Fact 5}

Similar to Fact 1, from Lemma \ref{03} we have
\begin{eqnarray}
&&|\langle E_{i}E_{j}E_{k}\rangle-\langle E_{i}\rangle\langle E_{j} E_{k}\rangle|
\nonumber\\
&\leq& ce^{-\kappa\tau}(e^{\kappa v t}-1) |X_{i}|\,|X_{j}|\,|X_{k}|
\label{eqapp100}
\end{eqnarray}
where $i,j,k\in\{1,\cdots,n\}$. With this result, we show this fact by induction.

The case of $k=3$ is followed from Eq.(\ref{eqapp100}). Assume that for $k=m<n$ we have
\begin{eqnarray}
 \nonumber&&|\langle E_{i_{1}}\cdots E_{i_{m}}\rangle-\langle E_{i_{1}}\rangle\cdots\langle E_{i_{m-2}}\rangle\langle E_{i_{m-1}}E_{i_{m}}\rangle|
 \\&\leq& ce^{-\kappa\tau}(e^{\kappa v t}-1)\Sigma^{m-2}_{i=1}\prod_{j=i}^m|X_{j}|
\label{eqapp101}
\end{eqnarray}

Define $\hat{\alpha}=\Sigma^{m-2}_{i=1}\prod_{j=i}^{m+1}|X_{j}|$. For $k=m+1$, we get that
\begin{eqnarray}
\langle E_{i_{1}}\cdots E_{i_{m+1}}\rangle
 &\leq&\langle E_{i_{1}}\rangle\cdots\langle E_{i_{m-2}}\rangle\langle E_{i_{m-1}}E_{i_{m}}E_{i_{m+1}}\rangle
   \nonumber\\
 &&
 +\hat{\alpha}ce^{-\kappa\tau}(e^{\kappa v t}-1)
     \nonumber\\
&\leq&\langle E_{i_{1}}\rangle\cdots\langle E_{i_{m-2}}\rangle(\langle E_{i_{m-1}}\rangle\langle E_{i_{m}}E_{i_{m+1}}\rangle
  \nonumber\\
 &&+ce^{-\kappa\tau}(e^{\kappa v t}-1) |X_{i_{m-1}}|\,|X_{i_{m}}|\,|X_{i_{m+1}}|)
   \nonumber\\
 &&+\hat{\alpha}ce^{-\kappa\tau}(e^{\kappa v t}-1)
  \nonumber\\
 &\leq&\langle E_{i_{1}}\rangle\cdots\langle E_{i_{m-1}}\rangle\langle E_{i_{m}}E_{i_{m+1}}\rangle
  \nonumber\\
 &&+ce^{-\kappa\tau}(e^{\kappa v t}-1)\Sigma^{m-1}_{i=1}\prod_{j=i}^{m+1}|X_{j}|
\label{eqapp102}
\end{eqnarray}
and
\begin{eqnarray}
\nonumber \langle E_{i_{1}}\cdots{}E_{i_{m+1}}\rangle
 &\geq&\langle E_{i_{1}}\rangle\cdots\langle E_{i_{m-2}}\rangle\langle E_{i_{m-1}}E_{i_{m}}E_{i_{m+1}}\rangle
 \\ \nonumber
 &&-\hat{\alpha}ce^{-\kappa\tau}(e^{\kappa v t}-1)
 \\ \nonumber
 &\geq&\langle E_{i_{1}}\rangle\cdots\langle E_{i_{m-2}}\rangle(\langle E_{i_{m-1}}\rangle\langle E_{i_{m}}E_{i_{m+1}}\rangle
 \\ \nonumber
 &&-ce^{-\kappa\tau}(e^{\kappa v t}-1) |X_{i_{m-1}}|\,|X_{i_{m}}|\,|X_{i_{m+1}}|)
  \\ \nonumber
 &&-\hat{\alpha}ce^{-\kappa\tau}(e^{\kappa v t}-1)
 \\
 \nonumber
 &\geq&\langle E_{i_{1}}\rangle\cdots\langle E_{i_{m-1}}\rangle\langle E_{i_{m}}E_{i_{m+1}}\rangle
 \\
 &&-ce^{-\kappa\tau}(e^{\kappa v t}-1)\Sigma^{m-1}_{i=1}\prod_{j=i}^{m+1}|X_{j}|
\label{eqapp103}
\end{eqnarray}
From the inequalities (\ref{eqapp102}) and (\ref{eqapp103}), we get that
\begin{eqnarray}
 \nonumber
 &&-ce^{-\kappa\tau}(e^{\kappa v t}-1)\Sigma^{m-1}_{i=1}\prod_{j=i}^{m+1}|X_{j}|
 \\ \nonumber
 &\leq&\langle E_{i_{1}}\cdots{} E_{i_{m+1}}\rangle-\langle E_{i_{1}}\rangle\cdots\langle E_{i_{m-1}}\rangle\langle E_{i_{m}}E_{i_{m+1}}\rangle
 \\
 &\leq& ce^{-\kappa\tau}(e^{\kappa v t}-1)\Sigma^{m-1}_{i=1}\prod_{j=i}^{m+1}|X_{j}|
\label{eqapp104}
\end{eqnarray}
So, from the inequality (\ref{eqapp104}) it follows that
\begin{eqnarray}
 \nonumber
 &&|\langle E_{i_{1}}\cdots{} E_{i_{s}}\rangle-\langle E_{i_{1}}\rangle\cdots\langle E_{i_{s-2}}\rangle\langle E_{i_{s-1}}E_{i_{s}}\rangle|
 \\ \nonumber
 &\leq& ce^{-\kappa\tau}(e^{\kappa v t}-1)\Sigma^{s-2}_{i=1}(|X_{i}|\,|X_{i+1}|\cdot\cdot\cdot|X_{s}|)
 \\
 &\leq& \alpha ce^{-\kappa\tau}(e^{\kappa v t}-1)
\label{eqapp105}
\end{eqnarray}

\end{document}